\documentclass[letterpaper,11pt]{article}
\usepackage{jheppub}
\pdfoutput=1
\usepackage{amsfonts}
\usepackage{amsmath}
\usepackage{amssymb}
\usepackage{comment}
\usepackage{float}
\usepackage{graphicx} 
\usepackage{braket}

\graphicspath{{./images/}}

\preprint{}
\title{Complex BPS Black Holes in AdS$_3\times S^3$}
\author{Finn Larsen and Kartik Sharma}
\affiliation{Leinweber Institute for Theoretical Physics, University of Michigan, Ann Arbor, MI, USA 48109}

\emailAdd{larsenf@umich.edu}
\emailAdd{kartiksh@umich.edu}

\date{April 2026}

\abstract{The correct gravitational representation of the supersymmetric index is a smooth complex solution, rather than the na\"{\i}ve Euclidean continuation of the Lorentizan BPS black hole. We construct such saddles for black holes with \(\mathrm{AdS}_3\times S^3\) boundary conditions in the STU model. We arrive at
the same finite-temperature BTZ \(\times S^3\) geometries in two independent ways: from two-center four-dimensional BPS solutions with complex dipole charges, and from six-dimensional black strings after imposing the BPS relation among conserved charges. We analyze the resulting smoothness and periodicity conditions and show that they are precisely those required for globally well-defined supersymmetry and for the thermodynamic potentials of the supersymmetric index.}

\begin{document}

\maketitle

\section{Introduction}

In quantum gravity, the leading contribution to the superconformal index \cite{Kinney:2005ej} is computed by a Euclidean saddle point
satisfying the appropriate asymptotic boundary conditions \cite{Cabo-Bizet:2018ehj,Iliesiu:2021are, Boruch:2023gfn, Sen:2007qy}.
The Euclidean continuation of the corresponding Lorentzian BPS black hole has an infinite throat and does not furnish a smooth finite-$\beta$ saddle with the index boundary conditions, so a more elaborate complex solution must be
identified. This strategy has been pursued successfully in many situations \cite{Boruch:2023gfn,Boruch:2025biv,Boruch:2025qdq,Boruch:2025sie,Cassani:2025iix,Adhikari:2024zif,Dharanipragada:2026dji,Chowdhury:2024ngg}, but the important case of black holes
with AdS$_3\times S^3$ boundary conditions was not yet studied from this point of view. The
purpose of this article is to fill this gap in the literature.

We pursue two complementary approaches to the problem which, in the end, lead to the same results.
First, we consider two center BPS black holes in four dimensions with complex dipole charges. Upon lifting to six
dimensions these black holes become finite temperature BTZ black holes multiplied by an $S^3$ that rotates with
complex velocity. Second, we consider general non-supersymmetric black strings in six dimensions and  impose the BPS condition relating the mass to other conserved charges. Once complex parameters are allowed for, we find finite temperature BTZ black holes again.

From both perspectives, the solutions have a simple interpretation. The na\"{\i}ve, singular BPS black hole is deformed by both thermal excitations and rotational motion. When complex parameters are taken into account their contribution to the energy can cancel, so that the BPS condition is satisfied. The regulated configurations satisfy smoothness conditions that can be exhibited in great detail, because this setting is very simple. The two ends of the dipole both represent $S^3$'s that shrink to zero size. Smoothness demands that the angles of both $S^3$'s satisfy specific periodicity conditions at their north and south poles. The condition that
all these local requirements  combine into a single consistent lattice is precisely such that globally well-defined Killing spinors are possible.

Let us explain our results in more detail. We work entirely in the STU model. The Bates--Denef multi-center solutions \cite{Bates:2003vx,Denef:2000nb,Denef:2007vg} in four dimensions depend on eight harmonic functions, corresponding to four electric and four magnetic charges. Following the standard 4D--5D connection, we obtain five-dimensional electric black holes by interpreting the magnetic charge \(P^0\) as the charge of a Taub--NUT geometry. The corresponding electric charge \(Q_0\) becomes angular momentum in five dimensions, and the resulting solution is the BMPV black hole \cite{Breckenridge:1996is}. Interpreting one of the three remaining electric charges as KK momentum, we lift the solution to a six-dimensional black string with an \(\mathrm{AdS}_3\times S^3\) region. The general form of supersymmetric solutions in five and six dimensions is known \cite{Gauntlett:2002nw,Gutowski:2003rg,Cariglia:2004kk}, and smooth horizonless solutions with these asymptotics underlie the microstate geometry program \cite{Bena:2004de,Bena:2005va,Bena:2007kg}.

These steps constitute a well known route that connects a black hole to a CFT$_2$ description \cite{Strominger:1996sh}. However, the Euclidean versions of the geometries involved behave poorly, as mentioned already. As a remedy, we follow the new attractor prescription \cite{Boruch:2023gfn}. Each harmonic function is divided into two centers, each carrying equal charge. Then imaginary dipoles, i.e. charges that cancel in pairs, are added, with values that render the entire configuration regular. The solution obtained in this way appears highly elaborate, but we show that it reduces to a BTZ $\times S^3$ black string that is completely standard, except that it features complex parameters.

We can reach the result in a complementary way, by temporarily disregarding supersymmetry and study general black holes in the 5D STU model. After lifting to six-dimensional black strings, the solutions depend on four conserved charges $(M, J, Q_R, Q_L)$ where $J$ refers to angular momentum within AdS$_3$ and the R-charges $Q_{R,L}$ are angular momenta on the $S^3$. In the grand-canonical ensemble, the solutions depend on the conjugate potentials $(\beta, \Omega, \Phi_R, \Phi_L)$. The explicit solutions are actually not presented in either set of thermodynamic variables, but in terms of auxiliary parameters $(m, \delta_3, l_1, l_2)$. We impose the BPS conditions from each of these perspectives and exhibit their relation.
For example, the unfamiliar relation between the parameters that appear in the solution
$$
\frac{2m}{\ell^2_3}e^{-2\delta_3} =\frac{2 (  l_1 +  l_2)}{\ell^2_3}e^{-\delta_3} -  1~,
$$
is equivalent to
$$
\frac{2\pi i}{\beta} =  2\Phi_L - \Omega - 1 ~.
$$
This thermodynamic form of the BPS constraint is similar to its analogues for higher-dimensional supersymmetric AdS black holes, whose solutions were constructed in \cite{Gutowski:2004yv,Chong:2005hr,Gutowski:2004ez} and whose constraints relating the chemical potentials were obtained in \cite{Hosseini:2017mds,Cabo-Bizet:2018ehj,Choi:2018hmj,Cassani:2019mms}.

This article is organized as follows.
In section \ref{sec:6Dsolutions} we start from the supersymmetric Bates-Denef two-center solutions in four dimensions with complex charges at the two centers specified according to the new attractor mechanism. We lift the solutions to six dimensions and show that the geometry is an $S^3$ fibered over BTZ.
In section \ref{sec:toBTZ} we study the solution and its parameters: the relation between the dimensionful supergravity charges and the quantized microscopic charges, the thermodynamics, and we detail the regularity conditions. In section \ref{sec:6Dstring} we start over again, from general non-supersymmetric black holes in five asymptotically flat dimensions. After lift to six dimensions and focus on a region that is asymptotically AdS$_3\times S^3$, we impose the BPS mass relation. We show that the results of our two independent computations agree after appropriate changes of variables.
We end, in section \ref{sec:discussion}, with a brief summary of our main results and an outlook towards future research.


\section{From 4D Multi-center to BTZ$\times S^3$}
\label{sec:6Dsolutions}

In this section we specify the four-dimensional two-center data in the Bates-Denef form and pick charges as dictated by the new attractor mechanism. We uplift to six dimensions and introduce ellipsoidal coordinates. We make it manifest that the resulting geometry is an $S^3$ fibered over the BTZ geometry.

\subsection{4D Data and its 6D Lift}

We begin by reviewing the chain of reinterpretations that leads from four-dimensional STU black holes to six-dimensional black strings with an asymptotic 
AdS$_3\times S^3$ region. 

The starting point is the stationary Bates--Denef multi-center BPS solution of the four-dimensional STU model. It is written in terms of eight harmonic functions on an auxiliary \(\mathbb R^3\),
\[
H^\Lambda=(H^0,H^1,H^2,H^3)~,
\qquad
H_\Lambda=(H_1,H_2,H_3,H_0)~,
\qquad
\Lambda=0,1,2,3 ~.
\]
The total conserved charge carried by these harmonic functions is the symplectic vector
\begin{equation}
\label{eqn:GamPQ}
\Gamma=(P^\Lambda,Q_\Lambda)~.
\end{equation}
The lift to five and six dimensions depends on this total charge vector and on the asymptotic values of the harmonic functions. It does not depend on whether the harmonic functions are represented by a single center or by a multi-center configuration with dipole terms. That distinction becomes important only later, when we impose regularity of the Euclidean saddle.

The standard 4D--5D lift identifies the magnetic charge \(P^0\) with the Kaluza--Klein monopole charge of a Taub--NUT space. The corresponding Gibbons--Hawking base is
\begin{equation}
\label{eqn:GHbase}
ds^2_{\rm GH}
=
\frac{1}{H^0}(dy+\chi)^2+H^0 d\vec x^{\,2},
\qquad
d\chi=*dH^0 .
\end{equation}
For the minimal quantized value of \(P^0\), the Taub--NUT geometry is locally \(\mathbb R^4\), written in Gibbons--Hawking coordinates. The four-dimensional electric charge \(Q_0\) becomes angular momentum in five dimensions, while the three electric charges
\[
Q_1~,\qquad Q_2~,\qquad Q_3~,
\]
become the three conserved electric charges of the five-dimensional STU black hole.

We will work in the mostly electric duality frame
\begin{equation}
    \label{eqn:Pi=0}
P^1=P^2=P^3=0 ~.
\end{equation}
This condition is important for the six-dimensional interpretation. It leaves \(P^0\) as the only magnetic charge, so that the five-dimensional solution is purely electric on a Taub--NUT base. One of the three five-dimensional electric charges can then be interpreted as Kaluza--Klein momentum along a compact circle, denoted \(z\). In the D1/D5 duality frame, \(H^0\) is the gravitational monopole background, \(H_1\) and \(H_2\) carry the two large D1/D5 charges, and \(H_3\) carries momentum along the D1/D5 intersection. Keeping the charge \(Q_0\) gives the BMPV generalization of the Strominger--Vafa black hole.

With these conventions the six-dimensional Lorentzian geometry becomes
\begin{equation}
\label{eqn:6Dmetric}
ds^2_6
=
-\frac{1}{\sqrt{Z_1Z_2}\,Z_3}(dt+\omega)^2
+\sqrt{Z_1Z_2}\,ds^2_{\rm GH}
+\frac{Z_3}{\sqrt{Z_1Z_2}}(dz-A^3)^2 ~,
\end{equation}
where
\begin{align}
Z_1
&=
H_1+\frac{H^2H^3}{H^0}~,
\qquad
Z_2
=
H_2+\frac{H^1H^3}{H^0}~,
\qquad
Z_3
=
H_3+\frac{H^1H^2}{H^0}~,
\\
\omega
&=
\omega_5(dy+\chi)+\hat\omega ~,
\\
*d\hat\omega
&=
\frac12\left(
H^0dH_0-H_0dH^0
+H^I dH_I-H_I dH^I
\right)~,
\label{eqn:omegahatdef}
\\
\omega_5
&=
\frac12 H_0
+\frac{H^IH_I}{2H^0}
+\frac{H^1H^2H^3}{(H^0)^2}~,
\\
\label{eqn:AIsoln}
A^I
&=
-\frac{1}{Z_I}(dt+\omega)
+\frac{H^I}{H^0}(dy+\chi)
-\xi^I~ , \qquad I=1,2,3 ~ ,
\\
*d\xi^I&=dH^I~ .
\end{align}
The Hodge star in these equations is the one on the auxiliary \(\mathbb R^3\). We keep the four-dimensional notation \(H^\Lambda,H_\Lambda\) after the lift because it is the most compact way to display the six-dimensional solution. The same formulae apply to single-center and multi-center configurations. The geometry \eqref{eqn:6Dmetric} is presented in Lorentzian signature. We primarily study the Euclidean continuation obtained by taking $t=-i\tau$ with $\tau$ real. We often present the geometries in their Lorentzian form but this continuation is implied. 

In six dimensions, the bosonic matter of the STU model becomes a scalar, a self-dual 2-form and an anti-self-dual 2-form. In the solution with metric \eqref{eqn:6Dmetric} they are also expressed in terms of the harmonic functions. The scalar has the form
\begin{align}\label{scalar6D}
   e^{X}= \frac{Z_2}{Z_1}~.
\end{align}
The self-dual 2-form and anti-self-dual 2-form fields combine into a single unconstrained 2-form $B$ with three-form field strength 
\begin{align}\label{6dGaugeform}
    G &= \left( \frac{Z_1^2}{Z_2 Z_3}\right)^{2/3}
\,\star_5 F^1
+
F^2\wedge(dz-A^3) ~.
\end{align}
The field strength $F^I = dA^I$ where $A^I$ was given in \eqref{eqn:AIsoln}. The Hodge star $\star_5$ is on the 5D reduction of the 6D metric. 

The asymptotic \(\mathrm{AdS}_3\times S^3\) boundary conditions are obtained by taking: 
\begin{equation}
\label{eqn:Hasymp}
H^0\sim \frac{P^0}{|\vec x|}~,
\qquad
H_{1,2}\sim \frac{Q_{1,2}}{|\vec x|}~,
\qquad
H_3\to h_3+\frac{Q_3}{|\vec x|}~,
\qquad
H_0\sim \frac{Q_0}{|\vec x|}~.
\end{equation}
The absence of constant terms in \(H_1,H_2\) is the decoupling limit; with nonzero constants the metric would instead be asymptotically flat in six dimensions. 
Thus \(Q_1\) and \(Q_2\) set the scale of the throat. By contrast, \(H_3\) retains a nonzero constant term $H_3\to h_3$
so that \(Q_3\) is interpreted as momentum excitation along the six-dimensional string. 

The absence of a constant term in \(H_0\) is also part of the black-string boundary conditions. We choose gauges such that
\[
\omega\to 0~,\qquad A^3\to 0~,
\]
up to terms that vanish at infinity. For $A^3$, this means a term $\frac{1}{h_3}dt$ must be added to the expression \eqref{eqn:AIsoln}.

In the purely Lorentzian single-center black string, the remaining magnetic harmonic functions vanish,
\[
H^1=H^2=H^3=0~ .
\]
In the Euclidean saddle these functions will be nonzero, but only as dipoles. They therefore fall off faster than the total-charge harmonic functions and do not change the conserved charges measured at infinity.


\subsection{The Euclidean Dipole Deformation}

The asymptotic boundary conditions for conserved charges $(P^\Lambda, Q_\Lambda)$ can be satisfied by harmonic functions that have a single, simple pole where $(H^\Lambda, H_\Lambda)\sim (\frac{P^\Lambda}{|\vec{x}|}, \frac{Q_\Lambda}{|\vec{x}|})$. However, for the BPS geometries we consider, the resulting infinite throat near the origin $\vec{x}\sim 0$ becomes singular. 

The new attractor mechanism \cite{Boruch:2023gfn} posits that the correct smooth Euclidean saddle is obtained by replacing the single pole in the harmonic functions by two poles, a north pole and a south pole. The total charge measured at infinity is kept fixed by each of the poles carrying $1/2$ of the total charges. In addition, imaginary charges $\pm i\delta_\Lambda$ are added with opposite signs in the two centers. These dipoles are invisible in the total conserved charges, but they are essential for regularity of the Euclidean geometry. Altogether, the dipole deformation replaces the vector of total charges \eqref{eqn:GamPQ} as
\begin{equation}
\label{eqn:Gammadef}
\Gamma^\alpha = (P^\Lambda,Q_\Lambda)~ \rightarrow \left(\frac{P^{\Lambda}}{2}\pm i\delta^\Lambda, \frac{Q_{\Lambda}}{2}\pm i\delta_\Lambda \right)~.
\end{equation}
The new attractor mechanism \cite{Boruch:2023gfn} gives an algorithm 
that determines the values of the dipole charges, which we explain in the following. 

As usual, the scalar moduli are assembled in the normalized period vector \begin{equation}
\label{eqn:period}
    \Omega^\alpha = e^{\frac{1}{2}K} (X^\Lambda,F_\Lambda)~.
\end{equation} 
The $F_\Lambda = \partial_\Lambda F$ where, in the STU-model, the prepotential $F$ is:
\begin{align}
F= -\frac{X^1\,X^2\,X^3}{X^0}~.
\end{align} 
The K\"{a}hler potential $K$ ensures the normalization condition $i \langle \Omega, \bar\Omega\rangle=1$. The usual attractor equations give algebraic relations between the charges $\Gamma^\alpha$ and the scalar moduli $\Omega^\alpha$:
$$
\Gamma^\alpha = i \Big( {\bar Z} (\Gamma, \Omega_*) \Omega^\alpha 
- Z(\Gamma,\Omega_*) \bar\Omega^\alpha \Big) ~.
$$
The normalization factor is the spacetime central charge:
\begin{equation}
\label{eqn:Zdef}
Z(\Gamma,\Omega) = \langle \Gamma, \Omega \rangle 
= e^{\frac{K}{2}} \Big( P^\Lambda F_\Lambda - Q_\Lambda X^\Lambda \Big)~.
\end{equation}
The bar on $\bar\Omega_*$ refers to complex conjugate, and the star refers to the specific function of charges that applies when the attractor equation is satisfied. 

The physical scalars $z^I$ are ratios between the projective coordinates $X^I$ and $X^0$. At the attractor point, they have the explicit form 
 \begin{align}\label{moduliformula1726}
z_*^I &=\frac{X^I}{X^0} = \frac{i P^I+ \partial_{Q_I}\Sigma(\Gamma)}{iP^0+\partial_{Q_0}\Sigma(\Gamma)}~.
\end{align}
We define 
\begin{equation}
\label{eqn:Sigmedef}
 \Sigma(\Gamma)= \sqrt{I_4(\Gamma)}~,
\end{equation}
where the quartic invariant: 
\begin{align}
    I_4(\Gamma)=-(P^\Lambda Q_\Lambda)^2+ 4(P^0Q_1Q_2Q_3-Q_0P^1P^2P^3+P^1Q_1P^2Q_2+P^2Q_2P^3Q_3+P^1Q_1P^3Q_3)~.
\end{align}
For an $8$ charge single center geometry at the attractor point, the central charge \eqref{eqn:Zdef} is simply
$$
Z_*(\Gamma) = i \Big(I_4(\Gamma)\Big)^{\frac{1}{4}} ~,
$$
when the arbitrary scale $X^0$ is chosen to be real.

So far, we only referred to a total charge vector which is real. Given this total charge, the complex dipole charges are generally specified as: 
\begin{align}
\label{deltacharge}
    \delta^\alpha = \frac{1}{2}\Big(\bar{Z}_{*}(\Gamma)\Omega^\alpha_{*}(\Gamma)+Z_{*}(\Gamma)\bar{\Omega}^\alpha_{*}(\Gamma)\Big)~.
\end{align}
After evaluating the period vector \eqref{eqn:period} in the STU-model, the dipole charges become\cite{Boruch:2025qdq}: 
\begin{align}
\label{deltachargeexplicit}
(\delta^0,\delta^I,\delta_I, \delta_0)= \left(-\frac{1}{2}\frac{\partial\Sigma(\Gamma)}{\partial Q_0},-\frac{1}{2}\frac{\partial\Sigma(\Gamma)}{\partial Q_I},\frac{1}{2}\frac{\partial\Sigma(\Gamma)}{\partial P^I}, \frac{1}{2}\frac{\partial\Sigma(\Gamma)}{\partial P^0}\right)~.
\end{align}

In our explicit computations we focus on the black string where magnetic charges vanish \eqref{eqn:Pi=0}. For this charge sector, the dipole charges \eqref{deltacharge} evaluate to: 
\begin{align}
    \delta^\alpha= \begin{pmatrix}
        \delta^0\\[6pt]
        \delta^1\\[4pt]
        \delta^2\\[4pt]
        \delta^3\\[4pt]
        \delta_1\\[4pt]
        \delta_2\\[4pt]
        \delta_3\\[4pt]
        \delta_0
    \end{pmatrix}  = \frac{1}{2\Big( 4P^0Q_1Q_2Q_3-(P^0Q_0)^2 \Big)^{\frac{1}{2}}} \begin{pmatrix}
(P^0)^2Q_0 \\[6pt]
-2P^0Q_2Q_3 \\[4pt]
-2P^0Q_1Q_3 \\[4pt]
-2P^0Q_1Q_2 \\[4pt]
-P^0Q_0Q_1 \\[4pt]
-P^0Q_0Q_2 \\[4pt]
-P^0Q_0Q_3 \\[4pt]
2Q_1Q_2Q_3-P^0Q_0^2
\end{pmatrix}~.
\end{align}
The harmonic functions $(H^0,H^I,H_I,H_0)$ with two centers that have complex charges specified in \eqref{eqn:Gammadef} now become: 
\begin{align}
\label{eqn:D1D5harm}
H^0&=\frac{P^0}{2}\left(\frac1{\rho_N}+\frac1{\rho_S}\right)
+i\,\frac{(P^0)^2Q_0}{2\Sigma}\left(\frac1{\rho_N}-\frac1{\rho_S}\right)~,\cr
H_I&=h_I+\frac{Q_I}{2}\left(\frac1{\rho_N}+\frac1{\rho_S}\right)
-i\,\frac{P^0Q_0Q_I}{2\Sigma}\left(\frac1{\rho_N}-\frac1{\rho_S}\right)~,
\cr
H_0&=\frac{Q_0}{2}\left(\frac1{\rho_N}+\frac1{\rho_S}\right)
+i\,\frac{2Q_1Q_2Q_3-P^0Q_0^2}{2\Sigma}\left(\frac1{\rho_N}-\frac1{\rho_S}\right)~,
\cr
H^I&=-\;i\,\frac{P^0Q_1Q_2Q_3}{Q_I\Sigma}\left(\frac1{\rho_N}-\frac1{\rho_S}\right)~.
\end{align}
Here \(\rho_N\) and \(\rho_S\) are the distances from the north and south centers in the auxiliary \(\mathbb R^3\). In each harmonic function the $\frac{1}{2}\Big(\frac{1}{\rho_N} + \frac{1}{\rho_S}\Big)$ term carries the total asymptotic charge. The charges $(P^\Lambda, Q_\Lambda)$ have unit of length and normalization as the pole of \(1/|\vec x|\). 
The $\frac{1}{\rho_N} - \frac{1}{\rho_S}$ term is a dipole that vanishes faster than \(1/|\vec x|\) at infinity, and therefore it does not contribute to the conserved charge. 

We ensure asymptotic AdS$_3\times S^3$ behavior by taking 
\[
h_1=h_2=0~
\]
as explained after \eqref{eqn:Hasymp}. In the mostly electric duality frame 
the ``spatial" magnetic charges vanish \eqref{eqn:Pi=0} and the function $\Sigma(\Gamma)$ defined in \eqref{eqn:Sigmedef} simplifies to
\begin{equation}
\label{eqn:sigmadef2}
\Sigma = \sqrt{4P^0 Q_1 Q_2 Q_3 - (P^0 Q_0)^2}~.
\end{equation}
The full six-dimensional geometry we study is \eqref{eqn:6Dmetric} with the $8$ harmonic functions specified in \eqref{eqn:D1D5harm}. This geometry should be smooth after appropriate periodicity conditions are imposed. Our detailed regularity analysis in section \eqref{periodicity} will validate this expectation.

\subsection{Ellipsoidal Coordinates}

For explicit computations, we employ ellipsoidal coordinates:
\begin{eqnarray}
x_1& = & a \sinh\gamma\sin\theta\cos\phi~,\cr
x_2& = &  a \sinh\gamma\sin\theta\sin\phi~,\cr
x_3 & =& a\cosh\gamma\cos\theta~.
\end{eqnarray}
For large $\gamma$, these coordinates reduce to spherical coordinates with radial coordinate $r = a\cosh\gamma$. The ellipsoidal coordinates focus attention on 
two poles 
$\vec{x}_{N,S}  =  \big( 0, 0, \pm a\big)$.
They are located at $\gamma=0$ and have $\theta=0$ and $\theta=\pi$, respectively. The distances to the North and South poles are
\begin{eqnarray}
\label{eqn:rhoNrhoSdef}
\rho_N & = & a ( \cosh\gamma - \cos\theta) ~,\cr
\rho_S & = & a( \cosh\gamma +  \cos\theta) ~.
\end{eqnarray}

In ellipsoidal coordinates, the Hodge duals of basic one-forms are
\begin{eqnarray}
\label{eqn:hodgedual}
*d\gamma &=& a\sinh\gamma\sin\theta  d\theta d\phi~,\cr 
*d\theta &=& a \sinh\gamma\sin\theta d\phi d\gamma~,\cr 
*d\phi &=& a\frac{\cosh^2\gamma - \cos^2\theta}{ \sinh\gamma\sin\theta }d\gamma d\theta~,
\end{eqnarray}
with orientation $(\gamma,\theta,\phi)$. We often need duals of harmonics that are centered at the poles:
\begin{eqnarray}
\label{eqn:dualrhon2}
  *d\frac{1}{\rho_N} & = & a  d\Big( \frac{\cosh\gamma\cos\theta-1}{\rho_N} d\phi \Big)~,\cr
*d\frac{1}{\rho_S} & = & a  d\Big( \frac{ \cosh\gamma\cos\theta+1}{\rho_S} d\phi \Big) ~.
\end{eqnarray}
We also have the Gibbons-Hawking identity: 
\begin{equation}
\label{eqn:dualcombo2}
 *\Big(  \frac{1}{\rho_N} d\frac{1}{\rho_S} - \frac{1}{\rho_S} d\frac{1}{\rho_N} \Big) = a  d \left(\frac{\sin^2\theta d\phi}{\rho_N \rho_S}\right) ~.
 \end{equation}
 
\subsection{The 6D Metric}

In this subsection, we make the 6D solution \eqref{eqn:6Dmetric} explicit. 
We start by computing the ingredients of the solution one by one. 
We first have: 
\begin{eqnarray}
\label{eqn:Z1expl}
Z_{1,2} & =&  H_{1,2} + \frac{H^{2,1} H^3}{H^0} =   \frac{Q_{1,2}}{aD} ~,\cr
Z_3&=&
H_3+\frac{H^1H^2}{H^0}=h_3+
\frac{Q_3}{
aD
} ~,
\end{eqnarray}
where 
\[
D=\cosh\gamma+i\frac{P^0Q_0}{\Sigma}\cos\theta~.
\]
The 6D scalar is proportional to the ratio $\frac{Z_1}{Z_2}$ so it is {\it independent} of the position. Next, we have: 
\begin{eqnarray}
\omega_5 & = & \frac{1}{2} H_0 + \frac{H^I H_I}{2H^0} +  \frac{H^1 H^2 H^3}{(H^0)^2} ~, \cr
&= &
\frac{Q_0\cosh\gamma}{
2aD^2
}
-
\frac{i\cos\theta}{D}\left[
\frac{h_3Q_1Q_2}{
\Sigma
}+
\frac{\Sigma}{
2aP^0D
}
\right]~.
\end{eqnarray} 

The remaining fields require Hodge duals on the base space. This requires additional care. 
First, we need
\begin{eqnarray}
*d\xi^3 &=&  dH^3 = -\;i\,\frac{P^0Q_1Q_2}{\Sigma}d\left(\frac1{\rho_N}-\frac1{\rho_S}\right) ~,\cr
*d\chi &=&  dH^0 =\frac{P^0}{2}d\left(\frac1{\rho_N}+\frac1{\rho_S}\right)
+i\,\frac{(P^0)^2Q_0}{2\Sigma}d\left(\frac1{\rho_N}-\frac1{\rho_S}\right)~.
\end{eqnarray}
The Hodge duals \eqref{eqn:dualrhon2} give: 
\begin{eqnarray}
\xi^{3}
&=&
\frac{2iP^0Q_1Q_2}{\Sigma}
\frac{\cosh\gamma\,\sin^2\theta}
{\cosh^2\gamma-\cos^2\theta}
\,d\phi  ~,\cr
\chi&=&
P^0
\left[
\frac{\sinh^2\gamma\,\cos\theta}{\cosh^2\gamma-\cos^2\theta}
-
i\frac{P^0Q_0}{\Sigma}
\frac{\cosh\gamma\,\sin^2\theta}{\cosh^2\gamma-\cos^2\theta}
\right]d\phi~.
\end{eqnarray}

For $\hat\omega$, we multiply out the definition \eqref{eqn:omegahatdef} in terms of harmonic functions and find: 
\begin{eqnarray}
2*_3d\hat\omega
=
\frac{iP^0Q_1Q_2h_3}{\Sigma}
d\left(\frac1{\rho_N}-\frac1{\rho_S}\right)
-
i\Sigma
\left(
\frac1{\rho_N}d\frac1{\rho_S}
-
\frac1{\rho_S}d\frac1{\rho_N}
\right) ~.
\end{eqnarray}
Then \eqref{eqn:dualrhon2} and \eqref{eqn:dualcombo2} give: 
\begin{eqnarray}
\label{eqn:rotpot2}
\hat\omega
=
-i\,\frac{\sin^2\theta}{\cosh^2\gamma-\cos^2\theta}
\left(
\frac{\Sigma}{2a}
+
\frac{h_3P^0Q_1Q_2}{\Sigma}\cosh\gamma
\right)d\phi ~.
\end{eqnarray}

We now have all contributions to the one forms that actually appear in the metric. In ellipsoidal coordinates, the GH base \eqref{eqn:GHbase} becomes
\begin{eqnarray}
\label{eqn:GHelliptic2}
\hspace{-1.cm}ds^2_{\rm GH}
&=&
\frac{a(\cosh^2\gamma-\cos^2\theta)}{P^0D}
(dy+\chi)^2
+
aP^0D
\left[
d\gamma^2+d\theta^2+
\frac{\sinh^2\gamma\,\sin^2\theta}
{\cosh^2\gamma-\cos^2\theta}
d\phi^2
\right]~,\cr
&=&
aP^0D\left(d\gamma^2+d\theta^2\right)\cr
&&+
\frac{a}{P^0}
\Bigg[
D\sin^2\theta\,dy^2
-
\frac{2iP^0Q_0}{\Sigma}\sin^2\theta\,dy\,\sigma^3_{P^0}
+
\frac{
\sinh^2\gamma-\frac{(P^0)^2Q_0^2}{\Sigma^2}\sin^2\theta
}{D}
(\sigma^3_{P^0})^2
\Bigg]
~, 
\end{eqnarray}
where $\sigma^3_{P^0}=\cos\theta\,dy+P^0d\phi $.
We also find
\begin{eqnarray}
\omega   &=&\omega_5 (dy + \chi) + \hat\omega \cr
&=&  \frac{Q_0\sin^2\theta}{2aD}\,dy
+
\frac{1}{D}
\left[
\frac{Q_0\cos\theta}{2a}
-
\frac{iQ_1Q_2}{\Sigma}
\left(h_3 + \frac{2Q_3}{aD}\right)
\right]\sigma^3_{P^0}~,
\end{eqnarray}
and
\begin{eqnarray}
A^3 &= & \frac{H^3}{H^0}  (dy + \chi) - \xi^3   - \frac{\omega}{Z_3}  + \Big(\frac{1}{h_3} - \frac{1}{Z_3} \Big)dt~, \cr
&=&
-\frac{Q_0\sin^2\theta}{2aDZ_3}\,dy
-
\frac{1}{DZ_3}\left[
\frac{Q_0\cos\theta}{2a}
+
\frac{i h_3Q_1Q_2}{\Sigma}
\right]\sigma^3_{P^0} +
\frac{Q_3}{h_3\left(aD\,h_3+Q_3\right)}dt ~.
\end{eqnarray}

At this point, we have computed all ingredients of the 6D metric \eqref{eqn:6Dmetric}, but the assembly is not entirely trivial. 
Some terms enjoy simplifications:
\begin{eqnarray}
&& - \frac{1}{Z_3}(dt+\omega)^2+ Z_3 \Big(dz  - A^3\Big)^2 
=Z_3\left(\frac{dt}{h_3}-dz\right)^2
-
2\,dt\left(\frac{dt}{h_3}-dz\right)\cr
&& - \frac{1}{D}\left[ \frac{Q_0\sin^2\theta}{a}
\left(
\frac{dt}{h_3}-dz
\right)dy+\left[ \frac{Q_0\cos\theta}{a}
\left(
\frac{dt}{h_3}-dz
\right)- \frac{2iQ_1Q_2}{\Sigma  } \left( dt+h_3dz \right) \right] \sigma^3_{P^0} \right]\cr
&&+
\frac{2iQ_1Q_2}{a\Sigma D^2}
\left[
Q_0\sin^2\theta\,dy\,\sigma^3_{P^0}
+
\left(
Q_0\cos\theta
-
\frac{2iQ_1Q_2Q_3}{\Sigma D}
\right)
(\sigma^3_{P^0})^2
\right]~, 
\end{eqnarray}
because $\omega$ and $A^3$ have many terms that are similar and group together nicely. 
The final result for the 6D metric becomes 
\begin{eqnarray}
ds^2
&=&
P^0\sqrt{Q_1Q_2}\,d\gamma^2
+
\frac{aD}{\sqrt{Q_1Q_2}}
\left[
Z_3\left(\frac{dt}{h_3}-dz\right)^2
-
2dt
\left(\frac{dt}{h_3}-dz\right)
\right]\cr
&&-\frac{Q_0\sin^2\theta}{\sqrt{Q_1Q_2}}
\left(
\frac{dt}{h_3}-dz
\right)dy
\cr
&&-\frac{1}{\sqrt{Q_1Q_2}}
\Bigg[
Q_0\cos\theta
\left(
\frac{dt}{h_3}-dz
\right)
-
\frac{2iaQ_1Q_2}{\Sigma}
(dt+h_3dz)
\Bigg]\left( \cos \theta dy +P^0 d\phi \right)\cr
&&+\sqrt{Q_1Q_2}P^0\left(\,d\theta^2
+
\frac{1}{(P^0)^2}
\left[
\sin^2\theta\,dy^2
+
\left(\cos \theta dy +P^0 d\phi \right)^2
\right]\right)
\cr
&=&
P^0\sqrt{Q_1Q_2}\,d\gamma^2
+
\frac{\Sigma^2}{4P^0(Q_1Q_2)^{3/2}}
\left(
\frac{dt}{h_3}-dz
\right)^2\\
&&-
\frac{a\cosh\gamma}{\sqrt{Q_1Q_2}}
\left(
\frac{dt}{h_3}-dz
\right)
\left(
dt+h_3dz
\right)
+
\frac{a^2P^0\sqrt{Q_1Q_2}}{\Sigma^2}
\left(
dt+h_3dz
\right)^2 \cr
&&+P^0\sqrt{Q_1Q_2} \Bigg( d\theta^2
+
\frac{1}{(P^0)^2}\sin^2\theta
\left(
dy
-\frac{P^0Q_0}{2Q_1Q_2}
\left[
\frac{dt}{h_3}-dz
\right]
\right)^2 \cr
&&+
\frac{1}{(P^0)^2}
\left[
\cos\theta\,\left( dy
-\frac{P^0Q_0}{2Q_1Q_2}
\left[
\frac{dt}{h_3}-dz
\right]\right) +P^0d\phi
+
\frac{iaP^0}{\Sigma}
\left[
dt+h_3dz
\right]
\right]^2 \Bigg)~.
\end{eqnarray}

At intermediate stages of the computation we have kept the constant term in the harmonic $H_3$ arbitrary and denoted $h_3$. Henceforth, we will take the conventional value $h_3=1$ and then the metric reduces to 
\begin{eqnarray}
\label{eqn:final6D}
ds^2&=&
\frac{a}{\sqrt{Q_1Q_2}}\Bigg(
\cosh\gamma\left(
-dt^2+dz^2
\right)+
\frac{\Sigma^2}{4aP^0Q_1Q_2}
\left(
dt-dz
\right)^2
+\frac{aP^0 Q_1 Q_2}{\Sigma^2}
\left(
dt+dz
\right)^2\Bigg)
\cr
&& +P^0\sqrt{Q_1Q_2}
\Bigg(d\gamma^2 + 
d\theta^2
+
\frac{\sin^2\theta}{(P^0)^2}
\Big(
dy
-\frac{P^0Q_0}{2Q_1Q_2}
\big(
dt-dz
\big)
\Big)^2\cr
&&
+
\frac{1}{(P^0)^2}
\left[P^0d\phi
+
\cos\theta
\Big(
dy
-\frac{P^0Q_0}{2Q_1Q_2}
\big(
dt-dz
\big)
\Big)
+
\frac{iaP^0}{\Sigma}
\big(
dt+dz
\big)
\right]^2
\Bigg)~.
\end{eqnarray}
The terms in the first line form, together with $P^0\sqrt{Q_1 Q_2}d\gamma^2$, the standard BTZ black hole in AdS$_3$. The last two lines are $S^3$ except for an imaginary shift along the $dt+dz$ direction and a real shift along $dt-dz$ direction of AdS$_3$. 
These correspond to Wilson lines. The $\Sigma$ is the duality invariant combination of conserved charges defined in \eqref{eqn:sigmadef2}. 

It is also interesting to work out the matter that supports the complex black hole solution. The three-form fluxes turn out to be constant, with the values they have in pure AdS$_3\times S^3$, except for the corotation that is identified already in the geometry \eqref{eqn:final6D}. Since this is outside the main thrust of this paper we present details in Appendix \ref{app:Matter}.

\section{Deciphering the BTZ Black Hole}
\label{sec:toBTZ}

The goal of this section is to extract CFT$_2$ data from the geometry \eqref{eqn:final6D} and interpret it physically.

\subsection{Units, Normalizations, and all That}
In arriving at \eqref{eqn:final6D} we diligently applied the convention that conserved charges $(P^\Lambda, Q_\Lambda)$ denoted by capital letters 
are lengths. They are all introduced in terms of the asymptotic behavior $H \sim \frac{Q}{|\vec{x}|}$ on the spatial base $\mathbb{R}^3$, and the corresponding Hodge-dual on $\mathbb{R}^3$. The conversion from supergravity charges to CFT$_2$ data involves many dimensionful scales. In addition to the $5$ physical charges, the supergravity solutions depend on the dipole radius $a$, the Planck scale $G_4$ and the radii $R_5, R_6$ of the $y, z$ circles that we lift the 4D solution on. 

As a first step towards variables that are more familiar from the AdS/CFT correspondence, we introduce the common scale $\ell^2_3= 4P^0\sqrt{Q_1 Q_2}$ for AdS$_3$ and $S^3$, and we trade the ellipsoidal coordinate $\gamma$ for the Fefferman-Graham coordinate through $\rho = \frac{1}{2} \gamma \ell_3$. This gives: 
\begin{eqnarray}
\label{eqn:final6D2}
ds^2  & = &  \frac{4P^0 a}{\ell^2_3}\cosh \frac{2\rho}{\ell_3}(-dt^2 + dz^2) +  \frac{\Sigma^2}{Q_1 Q_2\ell^2_3}( dt-dz)^2 + \frac{4a^2(P^0)^2Q_1 Q_2}{\Sigma^2\ell^2_3}( dt+dz)^2  + 
d\rho^2\cr
 && +  \frac{1}{4}\ell^2_3 \Bigg(d\theta^2 + \sin^2\theta \Big(\frac{dy}{P^0} - \frac{Q_0}{2Q_1 Q_2}(dt-dz)\Big)^2  \cr
 &&\quad \quad + 
\Big[ d\phi + \cos\theta \Big(\frac{dy}{P^0} - \frac{Q_0}{2Q_1 Q_2}(dt-dz)\Big) +\frac{ia}{\Sigma}(dt+dz)\Big]^2\Bigg)~.
\end{eqnarray}
Next, we recall that the geometry of the gravitational monopole \eqref{eqn:GHbase} depends on a single intrinsic scale which we denote $R_5$. Upon introducing the radial coordinate on the base $r = \sqrt{2R_5|\vec{x}|}$, the angular coordinate  $\psi= \frac{2}{R_5}y$, and the dimensionless monopole charge $p^0 = \frac{2}{R^5}P^0$, we find that the
entire geometry becomes $\mathbb{R}^4/\mathbb{Z}_{p^0}$ where $p^0$ is integral. Concretely, the conical singularity is due to the fractional angle $\psi/p^0$ appearing throughout. The solutions we consider are only asymptotic to the gravitational monopole geometry so, in general, they are asymptotically {\it locally} Euclidean, referred to as ALE. We will always take the quantized charge $p^0=1$ so that the asymptotic geometry is precisely $\mathbb{R}^4$, but the dimensionful variables involved in doing so are significant. From a practical point of view, we take $\frac{dy}{P^0}\to d\psi$. This simplifies \eqref{eqn:final6D2} and the identification $P^0 = \frac{1}{2} p^0 R_5$ will be needed in its own right.

The STU model is useful, because it is flexible. In both 4D and 5D  it is somewhat universal in that it incorporates many common embeddings into higher dimensions. 
In Kaluza-Klein reduction of string theory from 10D to 4D, a convenient benchmark is to represent all supergravity charges with dimension length as $Q= \sqrt{2G_4} n$ where $n$ is quantized as an integer, including magnetic charges. With this rule the 4D area law $S= \frac{1}{4G_4}A$ yields the familiar formula $S=2\pi\sqrt{p^0 q_1 q_2 q_3}$, where dimensionless charges indicated by lowercase letters are quantized. This rule applies literally when all charges are due to D-branes that wrap cycles with dimension $2\pi \sqrt{\alpha'}$. The contribution from a D-brane that wraps a cycle with radius $R$ is penalized by the inverse factor $\frac{\sqrt{\alpha'}}{R}$ and those that fail to wrap a cycle are enhanced by $\frac{R}{\sqrt{\alpha'}}$.

On our context, the D1- and D5-branes both wrap the $z$-circle which has radius $R_6$. Using the normalization rule from the previous paragraph directly in 6D, 
the scale of the AdS$_3$ and $S^3$ geometries becomes:
\begin{equation}
\label{eqn:l23}
\ell^2_3 = 4P^0 \sqrt{Q_1 Q_2} = 2R_5 p^0 \cdot  \frac{1}{2\pi R_6} \sqrt{2G_6q^1 q^2} = \frac{R_5}{\pi R_6} \sqrt{2G_6} \cdot \sqrt{q^1 q^2} ~.
\end{equation}
We take $p^0=1$ to ensure an asymptotically flat geometry. The ratio $\frac{R_5}{R_6}$ is a modulus of the torus that compactifies 6D to 4D in asymptotically flat space. 
This modulus drops out in the decoupling limit, according to the attractor mechanism, so we will omit it. 
Then the AdS$_3\times S^3$ version of the Brown-Henneaux \cite{Brown:1986nw} formula gives
\begin{equation}
\label{eqn:BHcentral}
k = \frac{\ell_3}{4G_3} = 2\pi^2 \cdot \frac{\ell^4_3}{4G_6} = q_1 q_2~.
\end{equation}
We have traded the central charge $c$ for the $SU(2)$ level $k$ through $c = 6k$ to avoid inconvenient factors of $6$ in many places. The final result $k=q_1 q_2$ is the familiar value for the D1-D5 system. 

In this article, we do not study the flow of moduli in any detail so, for our purposes, it is convenient to give all moduli a nominal value. 
Then we can track normalizations and dimensions by the generic rule $Q= \sqrt{2G_4} n$. Generic moduli can be restored by rescaling of charges. Additionally, in the decoupling limit we can identify the scale $\sqrt{R_5 R_6}$ that is inherited from a torus at infinity with $\ell_3$, without any loss of generality. 

With these simplifications, consider a pair of charges that are dual under electromagnetic duality. The normalization \eqref{eqn:BHcentral}
gives:
$$
P^0 Q_0  = 2G_4 p^0 q_0 =\frac{\ell^4_3}{4kR^2_6}\cdot p^0q_0 ~. 
$$
and then reorganization gives the quantization condition: 
\begin{equation}
\label{eqn:Q0quant}
Q_0 = \frac{\ell_3}{2} \cdot \frac{q_0}{k}~.
\end{equation}
Duality invariance ensures that we similarly have
\begin{equation}
\label{eqn:Q3quant}
Q_3 = \frac{\ell_3}{2} \cdot \frac{q_3}{k}~.
\end{equation}
For the remaining charges the simplified moduli give $Q_1 =Q_2 = P^0 = \frac{1}{2}\ell_3$.

As a check on these rules we read off the position of the BTZ horizon
$r_+ = \sqrt{4P^0 Q_3 - (Q_0)^2}$ and find that it corresponds to 
black hole entropy with the familiar normalization: 
$$
S = \frac{2\pi r_+}{4G_3} = \frac{2\pi k }{\ell_3}\cdot \sqrt{4P^0 Q_3 - (Q_0)^2} = 
2\pi \sqrt{k q_3 - \frac{1}{4k}q_0^2}~.
$$
The coupling $G_3$ was eliminated in favor of $k$ using \eqref{eqn:BHcentral}.

After rewriting into microscopic units, the black hole geometry \eqref{eqn:final6D2} becomes
\begin{eqnarray}
\label{eqn:final6D3}
ds^2  & = &  \frac{2a}{\ell_3}\cosh \frac{2\rho}{\ell_3}(-dt^2 + dz^2) +  \frac{kq_3 - \frac{1}{4}q_0^2}{k^2}(dt - dz)^2 + \frac{k^2}{kq_3 - \frac{1}{4}q_0^2} \cdot \frac{a^2}{\ell^2_3}\cdot (dt + dz)^2  + 
d\rho^2\cr
 && +  \frac{1}{4}\ell^2_3 \Bigg(d\theta^2 + \sin^2\theta \Big(d\psi - \frac{q_0}{k\ell_3}(dt-dz)\Big)^2  \cr
 &&\quad \quad + 
\Big[ d\phi + \cos\theta \Big( d\psi  - \frac{q_0}{k\ell_3}(dt-dz)\Big) +\frac{2aik}{\ell^2_3\sqrt{kq_3 - \frac{1}{4}q_0^2}}\cdot (dt+dz)\Big]^2\Bigg)~.
\end{eqnarray}
The rewriting uses 
$$
\Sigma = \sqrt{4P^0 Q_1 Q_2 Q_3 - (P^0Q_0)^2} =  \frac{\ell^2_3}{2k} \sqrt{kq_3 - \frac{1}{4}q_0^2}~,
$$
repeatedly. In the remainder or this section we will give most expressions both in terms of dressed supergravity charges $(P^0, Q_\Lambda)$ and the quantized charges $(1, q_\Lambda)$.


\subsection{Conformal Weights and Black Hole Thermodynamics}

In Schwarzchild-type coordinates, the BTZ black hole takes the form
\begin{eqnarray}
\label{eqn:BTZ}
\hspace{-2.cm} ds^2_{\rm BTZ}  &=&- \frac{(r^2-r^2_+)(r^2-r^2_-)}{r^2\ell^2_3}dt^2 + \frac{r^2\ell^2_3}{(r^2-r^2_+)(r^2-r^2_-)}dr^2 + \frac{r^2}{\ell^2_3}(dz - \frac{r_+r_-}{r^2}dt)^2
\end{eqnarray}
where $r_\pm$ are the coordinate positions of the horizons. After the coordinate transformation $r^2 = r^2_+ \cosh^2\frac{\rho}{\ell_3} - r^2_- \sinh^2\frac{\rho}{\ell_3}$, the AdS$_3$ part of 
our geometry \eqref{eqn:final6D2} takes precisely this form. The coefficients of 
$\frac{1}{\ell^2_3} (dz\pm dt)^2$ give the identifications
\begin{eqnarray}
\label{eqn:rpm1}
\frac{1}{4} (r_+ + r_-)^2  &= &  4P^0 Q_3 - \frac{(P^0 Q_0)^2}{Q_1 Q_2}~,\cr
\frac{1}{4} (r_+ - r_-)^2 & = & \frac{4a^2(P^0)^2 Q_1 Q_2}{4P^0 Q_1 Q_2 Q_3 - (P^0 Q_0)^2}
 ~.
\end{eqnarray} 
When reading off these values of $r_\pm$ the leading term in \eqref{eqn:final6D2}, proportional to 
$\cosh \frac{2\rho}{\ell_3}$, is immaterial. The normalization of that term can be absorbed in a linear shift of the radial coordinate $\rho$. 

The AdS$_3$/CFT$_2$ correspondence relates the horizon positions \eqref{eqn:rpm1} to the mass $M$ and the angular momentum $J$ of the  AdS$_3$ black hole in, or more generally to the energy momentum tensor in the dual CFT$_2$. It is convenient to focus on the
dimensionless conformal weights which, according to the AdS$_3$/CFT$_2$ dictionary, follow after multiplication by the universal factor $\frac{k}{\ell^2_3}$. The result is simplest when it is expressed in terms of the quantized charges: 
\begin{eqnarray}
\label{eqn:hLhR}
h_R&=& 
\frac{k}{4\ell^2_3}(r_+ + r_-)^2 =  q_3 - \frac{q_0^2}{4k}~,\cr
h_L &=& 
\frac{k}{4\ell^2_3}(r_+ - r_-)^2 =  \frac{a^2}{\ell^2_3} \cdot \frac{k^2}{q_3- \frac{q_0^2}{4k}}~.
\end{eqnarray} 
The level $k$ is large for macroscopic black holes. The quantized charges of the black holes are similarly large $q_3 \sim q_0 \sim k$ and so are the conformal weights $h_{L, R}\sim k$. The scale of the dipole $a\sim\ell_3$, just like the horizon positions $r_\pm\sim\ell_3$.
The black hole entropy is simply
\begin{equation}
\label{eqn:bhent1}
S = 2\pi \Big( \sqrt{kh_L} + \sqrt{kh_R}\Big)
= 2\pi \Big( \sqrt{kq_3 - \frac{1}{4}q_0^2} + \frac{ak^2}{\ell_3} \frac{1}{\sqrt{kq_3 - \frac{1}{4}q_0^2}}\Big)~.
\end{equation}
The position of the horizons also give the chiral temperatures
\begin{eqnarray}
\label{eqn:EbetaRL}
\beta_R & = & \frac{2\pi\ell^2_3}{r_+ + r_-}  =  \pi\ell^2_3\frac{\sqrt{Q_1Q_2}}{\sqrt{4P^0 Q_1 Q_2 Q_3 - (P^0 Q_0)^2}} = \frac{\pi\ell_3 k }{\sqrt{kq_3- \frac{1}{4}q_0^2}}
~, \cr
\beta_L & = & \frac{2\pi\ell^2_3}{r_+ - r_-}  = 
\frac{\pi\ell^2_3}{a} \frac{\sqrt{4P^0 Q_1 Q_2 Q_3 - (P^0 Q_0)^2}}{P^0\sqrt{Q_1Q_2}} =\frac{\pi\ell^2_3}{ak} \cdot  \sqrt{kq_3- \frac{1}{4}q_0^2}~.
\end{eqnarray}
These formulae give:
\begin{eqnarray}
\label{eqn:EbetaRL2}
a  = \frac{\pi^2\ell^3_3}{\beta_R\beta_L}~,
\end{eqnarray}
for the dipole radius $a$. The gravitational index depends on $\beta_R$ while $a$ or $\beta_L$ are two distinct regulators that are equivalent via \eqref{eqn:EbetaRL2}.

The black hole entropy \eqref{eqn:bhent1} is: 
\begin{equation}
\label{eqn:bhent2}
S = \frac{2\pi^2 k\ell_3}{\beta_R} + \frac{2\pi^2 k\ell_3}{\beta_L}~,
\end{equation}
so we can interpret \eqref{eqn:EbetaRL} as the effective temperatures of the oscillators that are responsible for the entropy. In Lorentzian signature, the extremal limit takes $\beta_L\to\infty$ with $\beta_R$ finite. The fugacity $\beta_R$ is conjugate to the conformal weight $h_R$ that remains free in the BPS limit and so becomes responsible for the BPS entropy. The deformation that is allowed in Euclidean signature does not change the structure in the R-sector but it introduces a finite inverse temperature $\beta_L$.

The chiral temperatures $\beta_{L,R}$ are related to the overall temperature through $\beta_{L,R} = \beta(1 \pm \Omega)$ where $\Omega$ is the rotational velocity of the black hole within AdS$_3$. It takes the value
$$
\Omega =\frac{r_-}{r_+} = 
\frac{ q_3-\frac{q^2_0}{4k} - \frac{ak}{\ell_3}}{q_3-\frac{q^2_0}{4k}+ \frac{ak}{\ell_3} }>0~.
$$
We assume positive rotation $\Omega>0$ so we have the inequality: 
\begin{equation}
\label{eqn:posOm}
q_3 - \frac{q^2_0}{4k} > \frac{ak}{\ell_3} ~.
\end{equation}
This is not a mere convention: if this condition is violated the temperature hierarchy is $\beta_L>\beta_R$, corresponding to level inversion. This can happen only if a state is created far from equilibrium artificially, or sustained by an external source \cite{Larsen:2025jqo}. In this work we generally keep $a$ small and therefore \eqref{eqn:posOm} will be satisfied easily.

According to \eqref{eqn:EbetaRL}, the inverse temperature can be recast as 
$$
\beta = \beta_R \left( 1  + \frac{\pi\ell^2_3}{a} \frac{4P^0 Q_1 Q_2 Q_3 - (P^0 Q_0)^2}{(P^0)^2 Q_1Q_2}  \right) = \beta_R \left( 1 + \frac{\ell_3}{ak^2} (kq_3 - \frac{1}{4}q^2_0)\right)~.
$$
Real black hole entropy requires $kq_3 - \frac{1}{4}q^2_0>0$ and the dipole scale $a$ is nonnegative, so we have the inequality $\beta\geq \beta_R$. Therefore, the Euclidean solutions obey a maximum on their temperature. 
This was remarked on previously \cite{Boruch:2023gfn}, \cite{Cassani:2025iix}, but the physical significance was mysterious. In our context, the interpretation is clear. 
The Lorentzian BPS black holes have physical degrees of freedom that account for their entropy and give a finite contribution to the inverse temperature. 
The finite temperature of the Euclidean BPS black hole introduces an independent source of thermal fluctuations, but this cannot lower the total inverse temperature.

\subsection{The Electric Potentials}

The electric potentials that appear in the first law of black hole thermodynamics 
\begin{equation}
\label{eqn:1stlaw1}
T dS = dM - \Omega dJ - \tilde\Phi_R dQ_R - \tilde\Phi_L dQ_L~,    
\end{equation}
are defined in a rotating frame. The right and left moving terms that appear in the metric are related to electric potentials as $\frac{\tilde\Phi_R}{1-\Omega} (dt-dz) = \tilde\Phi_R dt$ because $dz = \Omega dt$. When this is taken into account, the shifts of the $S^3$ angles $\psi$ and $\phi$ in the 6D metric \eqref{eqn:final6D3} give: 
\begin{eqnarray}
\label{eqn:Wilsonl}
\frac{2\tilde\Phi_R}{1-\Omega} &= &  \frac{q_0}{k} ~,\cr
\frac{2\tilde\Phi_L}{1+\Omega} & = & - \frac{2aik}{\sqrt{kq_3 - \frac{1}{4}q_0^2}} ~.
\end{eqnarray}
These Wilson lines give an additional contribution to the conformal weights
\cite{Kraus:2006nb}: 
\footnote{The relation to the notation of \cite{Kraus:2006nb}: ${\cal A}_w=\frac{2\tilde\Phi_L}{1+\Omega}$ and ${\cal A}_{\bar w}=\frac{2\tilde\Phi_R}{1-\Omega}$.}
\begin{eqnarray}
\label{eqn:hLhRadd}
k\left(\frac{\tilde\Phi_R}{1-\Omega}\right)^2 & = & \frac{q^2_0}{4k}~, \cr
k \left(\frac{\tilde\Phi_L}{1+\Omega}\right)^2 & = &  -  \frac{a^2 k^3}{kq_3 - \frac{1}{4}q_0^2}~, 
\end{eqnarray}
These formulae can be interpreted as the change in energy due to a spectral flow. 

The $h_R$ and $h_L$ given in \eqref{eqn:hLhR} are the irreducible conformal weights; they correspond to the free thermal contribution and do not incorporate Wilson lines. After adding \eqref{eqn:hLhRadd}, the total conformal weights become: 
\begin{eqnarray}
\label{eqn:hLhR2}
h^{\rm tot}_R&=& \frac{1}{2} (M+J)\ell_3 =  q_3 ~.\cr
h^{\rm tot}_L &=& \frac{1}{2} (M-J)\ell_3 = 0 ~.
\end{eqnarray} 
These are very natural results. In the R-sector, it means the total energy-momentum
$q_3$ is divided into a thermal part given in \eqref{eqn:hLhR} that is responsible for the entropy, and the part given in \eqref{eqn:hLhRadd} that carries the R-charge $Q_R$. This is the structure familiar from the original work on the BMPV 
black hole \cite{Breckenridge:1996is}. Presently, the energy-momentum in the L-sector similarly has two contributions, from thermal excitations and from the R-charge $Q_L$, and supersymmetry requires that they cancel. This is possible because the Wilson line in the L-sector is purely imaginary. Therefore, the black hole effectively saturates the extremality bound $M\ell_3 = J$, even though there is a thermal contribution.

The inverse temperature $\beta_L$ from \eqref{eqn:EbetaRL} and the electric potential $\Phi_L$ from \eqref{eqn:Wilsonl} are related as:
\begin{equation}
\label{eqn:BPSsat2}
\frac{2\pi i\ell_3}{\beta_L} = -\frac{2\tilde\Phi_L}{1+\Omega} ~.
\end{equation}
This is the BPS relation expressed in the grand canonical ensemble. It is simple, and it is natural that the BPS condition imposes a correlation between boundary conditions in AdS$_3$ ($\beta_L$) and boundary conditions on the $S^3$ ($\tilde\Phi_L$). However, the precise equation \eqref{eqn:BPSsat2} does not immediately match expectations from CFT$_2$, and from BPS black holes in higher dimensional AdS spacetimes. In this subsection we have added tilde to the potentials $\tilde\Phi_{R,L}$ to distinguish them from the more conventional ones. Importantly, the apparent discrepancy is not genuine, it is due to a mismatch in conventions that is worth explaining.

In CFT$_2$ on the cylinder, the eigenvalues of the Virasoro generators $L_0$ and $\bar L_0$ vanish in the NSNS vacuum and they equal $\frac{k}{4}$ in the RR vacuum. The Hamiltonian is smaller by $\frac{k}{4}$ due to a contribution from the Casimir energy so the NSNS vacuum is formally a black hole with $M\ell_3=-1$ and the RR vacuum has assignment $M\ell_3=0$. 

In the NSNS-vacuum, fermions are antiperiodic around the cylinder. The AdS$_3$ fills in the cylinder so this corresponds to antiperiodicity of a fermion that rotates an entire $2\pi$ around its own axis. This is the correct behavior for a particle in the spinor representation. In the RR-vacuum, fermions are periodic around the cylinder so this cycle cannot be contractible. To ensure supersymmetry, a Wilson line with strength $\Phi_L=\Phi_R=1$ must be included. However, the electric potentials \eqref{eqn:Wilsonl} that we read off from the geometry do not include such contributions. They must be interpreted as the addition to the bare potentials that are already present already in the RR-vacuum. In the following subsection we develop this from a different point of view. 



\subsection{BPS Black Holes with Finite Temperature: the NSNS-Perspective}

The similarities between black holes in AdS$_3\times S^3$ and their higher dimensional analogues is clearest in the NSNS-description. 
In this language, unitary representations of the supersymmetry algebra satisfy the BPS inequality \cite{Larsen:2021wnu}\footnote{In thermodynamic formulae, we use the normalizations and R/L convention from \cite{Larsen:2021wnu}. However, the notation there is related to the one here through: $P=J$, $J_{L,R} = Q_{L,R}$, $\omega_{L,R} =\Phi_{L,R}$. Additionaly, we freely take $\ell_3=1$ when no confusion can arise.}
\begin{equation}
\label{eqn:NSBPS}
    E + \frac{1}{2} k  \geq Q_L + J~.
\end{equation}
The constant $\frac{1}{2}k$ is due to the Casimir energy. Standard high-temperature thermodynamics derived from the Cardy-limit relates the AdS$_3$ quantum numbers $(E,J)$ to their conjugate potentials $(\beta,\Omega)$ as  
\begin{eqnarray}
\label{eqn:EJpot}
\frac{1}{2} ( E \pm J) & =&  \frac{k}{\beta^2(1\mp\Omega)^2}\Big( \pi^2 + 
\beta^2\Phi^2_{R,L}\Big)~.
\end{eqnarray}
The $S^3$ quantum numbers $(Q_L,Q_R)$ are similarly related to their conjugate potentials $(\Phi_L,\Phi_R)$ as 
\begin{eqnarray}
\label{eqn:QLRpot}
Q_{L,R} & = & \frac{2k\Phi_{L,R}}{1\pm \Omega}~.
\end{eqnarray}
These relations let us rewrite the BPS inequality \eqref{eqn:NSBPS}
in the grand canonical ensemble: 
\begin{equation}
\label{eqn:BPSineq1}
\frac{4\pi^2}{\beta^2}  +  \left( 1 + \Omega - 2\Phi_L\right)^2 \geq 0~.
\end{equation}
For real parameters, BPS saturation imposes {\it two} conditions
\begin{equation}
\label{eqn:BPSlor}
\beta=\infty~,\quad 1 + \Omega = 2\Phi_L~.
\end{equation}
They correspond to the relations between conserved charges: 
\begin{equation}
E =J + \frac{1}{2} k ~,\quad Q_L = k~.
\end{equation}
Therefore real BPS black holes can be parametrized by the angular momentum $J$ on AdS$_3$ and the R-charge $Q_R$. Unitarity only imposes $0\leq Q_L \leq 2k$, but BPS black holes also satisfy the constraint $Q_L=k$. However, complex solutions can 
saturate \eqref{eqn:BPSineq1} while imposing only {\it one} condition. It is either
\begin{equation}
\label{eqn:BPSsat1}
 1 + \Omega - 2\Phi_L =\frac{2\pi i}{\beta} ~,
\end{equation}
or its complex conjugate. This generalizes the two conditions \eqref{eqn:BPSlor} required when solutions are real. This form of the constraint between fugacities imposed by supersymmetry in AdS$_3\times S^3$ is the obvious adaptation of the analogous constraints satisfied by BPS black hole in higher dimensional AdS.

\begin{equation}
\label{eqn:BPSrel4}
\frac{2\Phi_L}{1+\Omega} = 1 - \frac{2\pi i}{\beta_L} ~.
\end{equation}
For real AdS$_3$ parameters $\beta$, $\Omega$, the potential $\Phi_L$ is genuinely complex, it has nontrivial real and imaginary parts.

The BPS constraint on the potentials \eqref{eqn:BPSrel4} is similar to \eqref{eqn:BPSsat2}, but they are not identical. As discussed at the end of the previous subsection, the CFT potentials $\tilde\Phi_{L,R}$ and the supergravity potentials $\Phi_{L,R}$ are related by a constant shift. In this subsection we identified the precise version of this shift:
\begin{equation}
\label{eqn:pot+1}
\frac{2\Phi_{L,R}}{1\pm\Omega} = 1 + \frac{2\tilde\Phi_{L,R}}{1\pm\Omega} ~.
\end{equation}

\subsection{Periodicity Conditions along Killing Directions}\label{periodicity}

The periodicities of the isometries parametrized by coordinates $t, z, \phi, \psi$ are constrained by regularity of the geometry. Some of the resulting conditions are local, others are global. In this subsection we discuss them in detail. 

\subsubsection{BTZ}
We first consider periodicity conditions for the BTZ geometry without taking the $S^3$ into account.

The BTZ geometry \eqref{eqn:BTZ} can recast as:
\begin{eqnarray}
ds^2_3 &=& \frac{1}{\ell^2_3} \Bigg( \cosh^2\frac{\rho}{\ell_3}  (r_+ dz- r_- dt)^2  - \sinh^2\frac{\rho}{\ell_3}  (r_+ dt - r_- dz)^2\Bigg) + d\rho^2\cr
&=& \cosh^2\frac{\rho}{\ell_3} d\eta^2 +  \sinh^2\frac{\rho}{\ell_3} d\xi^2 +  d\rho^2
~,
\end{eqnarray} 
where the dimensionless coordinates $(\xi,\eta)$ are
\begin{eqnarray}
\label{eqn:xietadef}
\xi & = & \frac{i}{\ell^2_3} ( r_+dt - r_- dz) =  \frac{1}{\ell^2_3} ( r_+d\tau - i r_- dz)~,\cr
\eta & = & \frac{1}{\ell^2_3} (  r_+ dz  - r_- dt ) = \frac{1}{\ell^2_3} (  r_+ dz  + i  r_- d\tau ) ~.
\end{eqnarray} 
The Wick rotation to Euclidean space sets $t=-i\tau$. The geometry near infinity is $-dt^2 + dz^2 = d\tau^2+dz^2$ up to a conformal factor that plays no role in our considerations. 

\begin{eqnarray}
\Delta \xi & = &   \frac{ r^2_+ - r^2_-}{r_+\ell^2_3} \beta  ~,
\cr
\Delta\eta  & = &   \frac{ i ( r^2_+ - r^2_-)}{r_-\ell^2_3}  \beta   = \frac{i r_+}{r_-} \cdot  \Delta \xi  ~.
\end{eqnarray}
The ratio of the two periods gives the purely imaginary and dimensionless complex structure 
$$
\frac{\Delta \eta}{\Delta\xi} = 
\frac{ir_+}{ r_-} = \frac{i}{ \Omega}~.
$$
We do not refer to complex structure by a specific symbol. The concept is not central to our considerations and we already denoted the Euclidean time by $\tau$. 

In the black hole background, the first of the two cycles $(d\xi, d\eta)$ is contractible, while the second is finite. The local geometry near the origin $\rho=0$ is 
$d\rho^2 + \rho^2 d\xi^2$ so the period of $\xi$ must be precisely $2\pi$ (up to a sign), or else the local $\mathbb{R}^2$ has a conical singularity. This condition determines 
$$
 \beta =  \frac{2\pi \ell^2_3r_+}{r^2_+ - r^2_-}~.
$$
This is the correct temperature. It agrees with $\beta = \frac{1}{2}(\beta_R + \beta_L)$ where $\beta_{R, L}$ are given in \eqref{eqn:EbetaRL}.

\subsubsection{$S^3$}
We now consider periodicity conditions for $S^3$ without accounting for BTZ. 
The $S^3$ geometry is
\begin{eqnarray}
ds^2_3 & =&  \frac{1}{4} \Big( d\theta^2 + d\psi^2 + (d\phi + \cos\theta d\psi)^2 \Big)
\cr
& =&   \frac{1}{4}d\theta^2 +\cos^2 \frac{\theta}{2} \cdot  \frac{1}{4} (d\psi + d\phi)^2 +  \sin^2 \frac{\theta}{2}  \cdot  \frac{1}{4}  (d\psi - d\phi)^2~.
 \end{eqnarray}
In the first line $S^3$ is presented as the Hopf fibration, in the second it is written in toric form. In the latter, it is manifest that smoothness near $\theta=0$ 
requires periodicity of  $\phi-\psi$ with period $4\pi$. We can do anything to  $\phi+\psi$. 
Similarly, smoothness near $\theta=\pi$ requires periodicity of  $\phi+\psi$ with period $4\pi$, with the action on $\phi+\psi$ unspecified. 
Taken together, these conditions define a two-dimensional lattice. It has two independent generators. $(\phi,\psi)\sim (\phi+2\pi,\psi+2\pi)$ and $(\phi,\psi)\sim (\phi+2\pi,\psi-2\pi)$.
Many other generators define the same lattice, including $(\phi,\psi)\sim(\phi+4\pi,\psi)$ and $(\phi,\psi)\sim(\phi+6\pi,\psi+2\pi)$. Importantly, $(\phi,\psi)\sim(\phi+2\pi,\psi)$ is not among the identifications. 

\subsubsection{The Twist}

The full 6D geometry can be presented as
\begin{eqnarray}
\label{eqn:final6D5}
ds^2  & = & \frac{2\pi^2\ell^2_3}{\beta_R\beta_L} \cosh \frac{2\rho}{\ell_3}(-dt^2 + dz^2) +  \frac{\pi^2\ell^2_3}{\beta^2_R}(dt - dz)^2 +  \frac{\pi^2\ell^2_3}{\beta^2_L} (dt + dz)^2  + 
d\rho^2\cr
 && +  \frac{1}{4}\ell^2_3 \Bigg(d\theta^2 + \sin^2\theta \Big(d\psi - \frac{2\Phi_R}{1-\Omega}(dt-dz)\Big)^2  \cr
 &&\quad \quad + 
\Big[ d\phi + \cos\theta \Big( d\psi  - \frac{2\Phi_R}{1-\Omega}(dt-dz)\Big)  - \frac{2\Phi_L}{1 + \Omega}  (dt+dz)\Big]^2\Bigg)~,
\end{eqnarray}
where the potentials are given in \eqref{eqn:Wilsonl}. We want to show that this geometry is smooth. Specifically, it cannot have conical singularities. 

The $\psi$ coordinate is shifted by a term $\frac{2\Phi_R}{1-\Omega}(dt-dz)$ that
becomes entirely imaginary after continuation to Euclidean signature, since $\frac{2\Phi_R}{1-\Omega}$ is real. Because this term is imaginary, this boundary condition serves as a fugacity and plays no role in regularity of the geometry. For the same reason the $+1$'s \eqref{eqn:pot+1} in the potentials due to the Casimir energy do not matter for regularity.  However, the shift in the $\phi$ coordinate is real, because $\frac{2\Phi_L}{1+\Omega}$ is purely imaginary, and this term is crucial for the regularity conditions. 

The AdS$_3$ base sets the period of $t+z $ to $-i\beta_L$ which, because of \eqref{eqn:BPSsat2}, is equivalent to $\phi\to \phi + 2\pi$. This shift  is only $1/2$ of the periodic identification $\phi\to\phi + 4\pi$ needed to ensure smoothness of $S^3$ by itself. 
However, the smoothness condition we seek does not concern $S^3$ by itself, it is for $S^3$ fibered over BTZ. The required identifications 
\begin{eqnarray}
\label{eqn:periods}
&&(t+z,t-z,\phi,\psi)\sim (t+z-i\beta_L, t-z-i\beta_R, \phi-2\pi,\psi)~,\cr
~
&& (t+z,t-z,\phi,\psi)\sim (t+z,t-z, \phi+2\pi,\psi+2\pi)~,\cr~
&&(t+z,t-z,\phi,\psi)\sim (t+z,t-z,\phi+2\pi,\psi-2\pi)~,
\end{eqnarray}
mix the coordinates nontrivially, but they define a consistent lattice. 

In the 6D geometry, the shift $t+z\to t+z-i\beta_L$ around the origin of BTZ is not a symmetry. One circumnavigation along this circle does not return to the original location, like a helical staircase. 
In the geometry here two circumnavigations do return to the original location, there is a hole that identifies two floors up with the ground floor. This global feature of the Euclidean geometry does not introduce any local feature, and certainly not a singularity.

\subsubsection{The Shrinking $S^3$'s}

We have introduced the geometry as a black hole augmented by a rotating $S^3$. It is interesting to consider the entire Euclidean geometry without prejudice towards a particular interpretation. For this we can disregard the finite circle parametrized by the coordinate $\eta$ and the shift proportional to the potential $\Phi_R$. Thus we consider a 5D geometry that is locally a product of the
${\mathbb R}^2$
$$
d\rho^2 + \rho^2 d\xi^2~,
$$
and the $S^3$:
 $$
 \frac{1}{4}d\theta^2 +\cos^2 \frac{\theta}{2} \cdot  \frac{1}{4} (d\phi + d\xi + d\psi)^2 +  \sin^2 \frac{\theta}{2}  \cdot  \frac{1}{4}  (d\phi + d\xi - d\psi)^2~.
 $$
 The main subtlety is the  twist by $\xi$ in the azimuthal angle $\phi$ on $S^3$. However, in the following, we want to be careful about the geometry when $\rho=0$ and $\sin\theta=1$ simultaneously.

The full 5D geometry near $\theta=0$ is: 
\begin{eqnarray}
ds^2_5 &=& \frac{1}{4} \Big( d\theta^2 + \theta^2 \cdot \frac{1}{4}(d\phi+ d\xi -d\psi)^2 \Big)+ d\rho^2 + \rho^2 d\xi^2 + \frac{1}{4} (d\phi+d\xi+d\psi)^2 \cr
&=& dR^2 + R^2 \Big(\cos^2\alpha \cdot \frac{1}{4}(d\phi + d\xi -d\psi)^2 +  \sin^2\alpha d\xi^2\Big)+ \frac{1}{4} (d\phi+ d\xi +d\psi)^2~, 
\nonumber
\end{eqnarray}
Here $\rho = R\sin\alpha$ and $\theta = 2R\cos\alpha$. 
The analogous geometry near $\theta=\pi$ is: 
\begin{eqnarray}
ds^2_5 &=& \frac{1}{4} \Big( d\theta^2 + (\pi -\theta)^2 \cdot \frac{1}{4}(d\phi+d\xi +d\psi)^2  \Big)+ d\rho^2 + \rho^2 d\xi^2 ) + \frac{1}{4} (d\phi+d\xi -d\psi)^2 \cr
&=& dR^2 + R^2 \Big(\cos^2\alpha \cdot \frac{1}{4}(d\phi+ d\xi +d\psi)^2 +  \sin^2\alpha d\xi^2\Big)+ \frac{1}{4} (d\phi + d\xi-d\psi)^2~,
\nonumber
\end{eqnarray}
where $\rho = R\sin\alpha$ and $\pi-\theta = 2R\cos\alpha$. For both values of $\theta$, the local geometry is $\mathbb{R}^4$, as long as the $S^3$'s that contract to zero size at $R=0$ satisfy the correct periodicity conditions. That is needed to 
avoid conical singularities. 

Global regularity of the geometry imposes four conditions, arising as regularity conditions on the north and south poles of two $S^3$'s that shrink at separate loci. 
For both $\theta=0$ and $\theta=\pi$, the regularity near $\alpha=0$ requires $\xi$ having period $2\pi$, as expected on $\mathbb{R}^2$. Importantly, periodicity $2\pi$ at the two points must be taken with the distinct combinations $\phi + \xi \pm \psi$ fixed. Regularity near $\alpha=\frac{\pi}{2}$ similarly requires $\phi+\xi\pm\psi$ having period $4\pi$ for the two values of $\theta$. The geometry is regular as a whole because all four requirements are satisfied by the three periodicity conditions \eqref{eqn:periods}.


\subsubsection{Boundary Conditions on Fermions}
For supersymmetry, the preserved fermionic symmetry must be well defined throughout the manifold. This condition is distinct from regularity of the bosonic manifold and generally thought about as stronger. There must be a consistent spin structure. 

As a benchmark, consider a smooth 2D spacetime that caps off in the sense that it becomes locally $\mathbb{R}^2$ at some tip. 
A good example to have in mind is the Euclidean Schwarzchild black hole in its near horizon region. In that setting, a circle at fixed distance from the origin is contractible so, upon circumnavigation, it is equivalent to a complete rotation of a particle around its own axis. By the spin-statistics theorem, fermions transform in a $\frac{1}{2}$-integral representation of the local rotation group. These are not proper representations, in the sense that the field acquires a sign upon a complete rotation. Usually this is not a problem, it is the behavior of electrons studied in a first introduction to quantum mechanics. However, supersymmetry interchanges bosons and fermions so it can only be a symmetry if it preserves boundary conditions, and for that a fermion must have periodic boundary conditions. Supersymmetry is difficult to achieve because of the tension between boundary conditions on fermions that must be periodic under supersymmetry and antiperiodic under local rotations. 

In our conventions, the supersymmetry that is well-defined in AdS$_3\times S^3$ is a spinor in the $L$ sectors of both AdS$_3$ and $S^3$, but a scalar in the $R$-sector. In the geometry, the BTZ coordinate $t+z$ that corresponds to a contractible cycle is not periodic by itself, it is correlated with the angle $\psi$ on $S^3$:
\begin{equation}
\Psi(t+z-i\beta_L,\phi-2\pi) = \Psi(t+z,\phi)~.
\end{equation}
We have supressed $t-z$ and $\psi$ because, for the fermion, they are inert.
This is the global condition imposed by supersymmetry. These boundary conditions  do not depend on the radial position $r$, they are the same at the horizon and at infinity. They also do not depend on the polar angle $\theta$ on $S^3$. Moreover, the potential $\Phi_R$ does not enter, 
because the preserved supersymmetry is neutral in the R-sector.

\section{From General Nonextreme BH to BPS}
\label{sec:6Dstring}
In higher dimensional AdS spacetimes, Euclidean black holes that are BPS and have finite temperature can be identified by considering a general black hole and then imposing the BPS condition. For Lorentzian black holes the BPS condition can be satisfied only when an extremality condition is imposed as well, so finite temperature is precluded. However, this implication can be circumvented by complex black holes in Euclidean signature. In this section we 
carry out this logic for AdS$_3\times S^3$. \footnote{This subsection mostly follows \cite{Cvetic:1998xh} but notation is modernized.  $\lambda_{\rm there} = \ell_3= (Q_1 Q_2)^{\frac{1}{4}}$ and the indices $i=0,1,2 \to i=1, 2, 3$. We consistently interchanged $L\leftrightarrow R$. The potentials for R-charge $\Phi_{L,R}$ were denoted  $\Omega_{R,L}$ there. $y$ and $R_y$ are $z$ and $R_6$ here. 
The toric angles there $(\phi, \psi)$ are denoted $(\phi_1, \phi_2)$ here and should not be confused with our angles in Hopf notation. Finally, the angular momentum of the 5D black hole ,$j^{\rm here}_{R,L} = 2j^{\rm there}_{R,L}$.}

\subsection{Thermodynamics from Asymptotically Flat  5D BH's}
The black holes in STU supergravity with five asymptotically flat dimensions depend on the six physical variables mass $M$, two angular momenta $J_{L,R}$, and three charges $Q_i$. The analytical presentation of the solution involves a mass parameter $m$ with dimension length squared, two angular momenta parameters $l_{1,2}$ with dimension length, and three boost parameters $\delta_i$ that are dimensionless. The relation between the physical variables and the parameters that appear in the solution is: 
\begin{eqnarray}
\label{eqn:6Dstr}
M & = & \frac{\pi}{4G_5}\cdot m \sum_{I=1}^3 \cosh 2\delta_I ~, \cr
j_{R,L} & = &  \frac{\pi}{4G_5} \cdot 2m (l_1\mp l_2) (\prod_{I=1}^3 \cosh\delta_I \pm \prod_{I=1}^3 \sinh\delta_I )~,\cr
Q_I & = &  \frac{\pi}{4G_5} \cdot m \sinh 2\delta_I  ~, \quad I=1,2, 3~.
\end{eqnarray}
Lower-case variables like $j_{R,L}$ refer to quantum numbers. The notation $J_{R,L}$ refers to angular momenta and is natural in this section, but it is interchangeable with $Q_{R,L}$ that stresses rotation on the $S^3$ of AdS$_3\times S^3$. 
 The electric $Q_I$ here have dimension inverse length. The formulae for $Q_I$ with the normalization factor $\frac{\pi}{4G_5}$ omitted has dimension length squared and correspond to pole charges in 5D; they appear in a 5D harmonic function as $H=1+\frac{Q}{r^2}$.  Comparison of fluxes from harmonic functions in 5D and 4D gives the dictionary between pole charges: 
\begin{equation}
\label{eqn:4D5DQs}
Q_{\rm 5D} = 2R_{11} Q_{\rm 4D}~.
\end{equation}
Thus the dimensionful supergravity charges in this section are normalized differently from those in earlier sections, even though they are denoted by the same symbol. Apart from this normalization convention, the $Q_I$ that appear in this section can be identified with the charges introduced in \ref{eqn:GamPQ} and used throughout section \ref{sec:6Dsolutions}.

We interpret the 5D black holes as 6D black strings and then all the conserved charges \eqref{eqn:6Dstr} become uniform densities. More importantly, we take the decoupling limit $\delta_{1,2}\to\infty$. The total mass $M$ in \eqref{eqn:6Dstr} is attributed symmetrically to the three charges $Q_1$, $Q_2$, and $Q_3$ but, in the decoupling, $Q_{1,2}$ become parametrically large. We can then introduce an excitation energy $E$ that excludes their contribution and \eqref{eqn:6Dstr} become: 
\begin{eqnarray}
\label{eqn:AdSvariables}
\frac{E}{2\pi R_6} & = & \frac{k}{2\pi\ell^4_3} \cdot m  \cosh 2\delta_3 ~, \cr
\frac{Q_3}{2\pi R_6} & = &  \frac{k}{2\pi\ell^4_3}   \cdot m \sinh 2\delta_3~,\cr
\frac{ j_{R,L}}{2\pi R_6} & = &  \frac{k}{2\pi\ell^2_3}   \cdot  (l_1\mp l_2) e^{\pm\delta_3}  ~.
\end{eqnarray}
In the near horizon limit, we simplified the gravitational coupling as 
$$
\frac{\pi}{4G_5}  =  \frac{2\pi^2 R_6}{4G_6}  = \frac{2\pi^2 R_6}{4G_3\cdot 2\pi^2 \ell^3_3}  = \frac{k}{2\pi \ell^4_3}\cdot 2\pi R_6~.
$$
where the level of dual CFT$_2$ is \eqref{eqn:BHcentral}. The factor $2\pi R_6$ is highlighted because the conserved charges become densities after lift to 6D. 
The energy $E$ and the momentum charge $Q_3$ both have dimension inverse length. The dimensionless angular momentum quantum numbers $j_{L,R}$ are quantized. \footnote{The normalization of $j_{L,R}$ is such that it can be $\frac{1}{2}$-integral or integral. In Cartan conventions $SU(2)$ representations are labelled by integers, in analogy with $p$ for a particle in a box. The dictionary is $j_{\rm this section} = \frac{1}{2}j_{\rm Cartan}$.} In the same parametric notation, the potentials become:
\begin{eqnarray}
\label{eqn:5Dpotentials}
\beta^{R,L} & = &\beta_H (1 \mp \Omega) =  \frac{2\pi\ell^2_3 e^{\mp\delta_3}}{\sqrt{2m - (l_1\mp l_2)^2}} ~,
\cr
\beta_H \Phi^{R,L} & = & \frac{\pi (l_1\mp l_2)}{\sqrt{2m - (l_1\mp l_2)^2}} ~.
\end{eqnarray}
The inverse temperature $\beta_H$ is a length, $\Omega$ is dimensionless, and the electric potentials $\Phi_{R,L}$ have dimension inverse length. 

In the decoupling limit the $Q_1$ and $Q_2$ charges source the asymptotic AdS$_3\times S^3$ background and determine the scale of both factors as 
$$
\ell^2_3 = 2m \cosh\delta_1 \cosh\delta_2 =  \frac{4G_5}{\pi}  \cdot \sqrt{Q_1 Q_2}~.
$$
This expression agrees precisely with its analogue \eqref{eqn:l23} that applies to geometries derived from superposition of harmonic functions in 4D when the dictionary in \eqref{eqn:4D5DQs} is taken into account. Therefore, the familiar relation \eqref{eqn:BHcentral} follows from this formula, as it should. 

The central features of the complex Euclidean saddle points follow from basic thermodynamics, without considering the geometry explicitly. 

The energy $E$ and momentum charge $Q_3$ given in \eqref{eqn:AdSvariables} are quantized as $E=\frac{\epsilon}{R_6}$ and $Q_3=\frac{q_3}{R_6}$, respectively. The dimensionless parameters $\epsilon$ and $q_3$ are the integral quantum numbers of a particle in a box with periodic boundary conditions. 
Therefore, the effective conformal weights become: 
\begin{equation}
\label{eqn:hLRsugra}
h_{R,L} = \frac{1}{2} ( \epsilon \pm  q_3) - \frac{1}{4k}j^2_{R,L} = \frac{kR^2_6}{4\ell^4_3} \cdot  \Big( 2m -(l_1\mp l_2)^2\Big) e^{\pm2\delta_3}~.
\end{equation}
Later, the explicit geometry will show that the factor $e^{\pm2\delta_3}$ is a ``boost" factor along the 6th dimension that contains the distinction between total energy and momentum along the string. The overall factor distinguishes between the scale $R_6$ of a particle in a box and the AdS$_3$ scale $\ell_3$ but, because of conformal invariance, no loss of generality is lost by taking $R_6=\ell_3$. It is interesting that the angular momenta ``belong"  to the two chiralities in a manner that is not manifest from the 5D perspective.

\subsection{General Black Strings with Finite Temperature}

We now study the explicit geometry of the 6D black string in the near horizon limit \cite{Cvetic:1998xh}: 
\begin{eqnarray}
\label{eqn:6Dv2}
ds^2_6 & =&  - \frac{(r^2+r^2_1)(r^2+l^2_2)-2mr^2}{\ell^2_3r^2} d\tilde{t}^2 + \frac{r^2}{\ell^2_3} (d\tilde z - \frac{l_1 l_2}{r^2} d\tilde t)^2
\cr
& +&  \frac{\ell^2_3r^2} {(r^2+l^2_1)(r^2+l^2_2)-2mr^2}dr^2 + \ell^2_3\Big( d\Theta^2 + \sin^2\Theta d\tilde\phi^2_1  + \cos^2\Theta d\tilde\phi^2_2\Big)~.
\end{eqnarray}
The coordinates with tilde are boosted along the string: $\tilde t = t\cosh\delta_3 - z \sinh\delta_3$, $\tilde z = z\cosh\delta_3 - t \sinh\delta_3$ and the angular coordinates with tilde were shifted to a frame that is both rotating and boosted: 
\begin{eqnarray}
\tilde{\phi}_2 & = & \phi_2 - \frac{1}{\ell^2_3} (l_2 \tilde t + l_1 \tilde z) = \frac{\psi +\phi}{2}  - \frac{l_1 + l_2}{2\ell^2_3}e^{-\delta_3} (t + z) - \frac{l_1 - l_2}{2\ell^2_3} e^{\delta_3}  (t-z)  ~, \cr
\tilde{\phi}_1 & = & \phi_1 - \frac{1}{\ell^2_3} (l_1 \tilde t + l_2 \tilde z)
=  \frac{\psi -\phi}{2}   - \frac{l_1 + l_2}{2\ell^2_3}e^{-\delta_3} (t + z) + \frac{l_1 - l_2}{2\ell^2_3} e^{\delta_3} (t-z) ~.
\end{eqnarray}
Note that the angles in \eqref{eqn:6Dv2} and toric, not to be confused with the ones that appear when $S^3$ is presented as a Hopf fibration. The corresponding polar angle $\Theta=\frac{1}{2}\theta$. 

The implicit boosts are awkward but they can be unpackaged nicely. The horizon coordinates of the geometry presented as \eqref{eqn:6Dv2} 
are: 
\begin{eqnarray}
r^2_\pm 
& = & \frac{1}{4} \Big( \sqrt{ 2m - (l_1 +l_2)^2 } \pm  \sqrt{ 2m - (l_1 -l_2)^2}\Big)^2~.
\end{eqnarray}
After a simple shift in the radial coordinate to: 
$$
\bar r^2 = r^2  + \Big( 2m - l_1^2- l^2_2\Big) \sinh^2\delta_3 + 2l_1 l_2 \sinh\delta_3 \cosh\delta_3 ~,
$$
they become:
\begin{eqnarray}
{\bar r}^2_\pm 
& = & \frac{1}{4} \Big( \sqrt{ 2m - (l_1 +l_2)^2 } e^{-\delta_3} \pm  \sqrt{ 2m - (l_1 -l_2)^2}e^{\delta_3}\Big)^2~.
\end{eqnarray}
This is equivalent to: 
\begin{equation}
\label{eqn:rpmtemp}
(\bar r_+ \pm \bar r_- )^2 = \Big(2m - (l_1 \mp l_2)^2\Big)e^{\pm 2\delta_3}~.
\end{equation}
Thus the complicated looking linear transformations are entirely captured by the boost factors $e^{\pm 2\delta_3}$, as expected because the $(t,z)$ variables are string coordinates that enjoy Lorentz symmetry. Manipulation of the terms in the first line of the metric \eqref{eqn:6Dv2} confirms that the metric is in fact BTZ with these parameters. 

For reference, we write the entire metric in standard form: 
\begin{eqnarray}
\label{eqn:gen6Dfinal}
ds^2_6 & =&  - \frac{(\bar r^2-\bar r^2_+)(\bar r^2-\bar r^2_-)}{\ell^2_3\bar r^2} dt^2 + \frac{\bar r^2}{\ell^2_3} (dz - \frac{\bar r_+ \bar r_-}{\bar r^2} d\tilde t)^2 + 
 \frac{\ell^2_3\bar r^2}{(\bar r^2-\bar r^2_+)(\bar r^2-\bar r^2_-)} d\bar r^2
\cr
&+&
\frac{\ell_3^2}{4}
\Bigg[
d\theta^2
+\sin^2\theta
\left(
d\psi
-
\frac{l_1+l_2}{\ell_3^2}e^{-\delta_3}(dt+dz)
\right)^2
\cr
&+&
\Bigg(
d\phi
-
\frac{l_1-l_2}{\ell_3^2}e^{\delta_3}(dt-dz)
+\cos\theta
\left(
d\psi
-
\frac{l_1+l_2}{\ell_3^2}e^{-\delta_3}(dt+dz)
\right)
\Bigg)^2
\Bigg]~,
\end{eqnarray}
This is the near horizon geometry of the 6D black string, when presented as $S^3$ fibered over BTZ. This result has been recognized since \cite{Lunin:2002iz}.
Our objective is to apply the result when parameters are complex.

\subsection{BPS Condition in Supergravity Variables}

The final result \eqref{eqn:gen6Dfinal} of the previous subsection is a large family of 6D black strings that is generic, in the sense that supersymmetry plays no special role. In order to identify the subset of solutions that are supersymmetric we must discuss the BPS condition in the supergravity variables being used there. 

The BPS condition for a sector of CFT$_2$ with fixed conserved charges is \cite{Larsen:2021wnu} 
\begin{equation}
\label{eqn:BPSineq}
\epsilon \geq q_3 + j_L  -  \frac{1}{2} k~.
\end{equation}
It is significant that the equation is linear in charges, in contrast to the quadratic dependence on angular momentum that is typical for the conformal weight \eqref{eqn:hLRsugra}. Rewriting the latter using the BPS inequality \eqref{eqn:BPSineq}, we find 
\begin{equation}
\label{eqn:hLBPS}
h_L = \frac{1}{2}(\epsilon-q_3) - \frac{1}{4k}j^2_L \geq  - \frac{1}{4k}(j_L-k)^2~.
\end{equation}
For real black holes in Lorentzian signature the effective conformal weight $h_L\geq 0$ so the BPS bound can be saturated only if both $h_L=0$ {\it and} $j_L=k$. In the BPS limit there can be no thermal excitations {\it and} 
the angular momentum is required to take a specific nominal value. 

However, this assumes all parameters of the black hole are real. For complex solutions the BPS condition can be satisfied with values $h_L<0$ that are correlated with $j_L$ in the manner prescribed by saturation \eqref{eqn:hLBPS}. In the conventions inherited from 6D supergravity, the left hand side of \eqref{eqn:hLBPS} is
\begin{equation}
\label{eqn:hLsugra}
h_L = k \left( \frac{\pi R_6}{\beta_{L}}\right)^2 = \frac{k}{4\ell^2_3} \cdot  \Big( 2m - (l_1+ l_2)^2\Big) e^{-2\delta_3}~,
\end{equation}
and the right hand side gives: 
\begin{eqnarray}
\label{eqn:jLsugra}
- \frac{1}{4k} \Big( j_L  - k \Big)^2   & = & - \frac{1}{4k} \Big( \frac{\pi}{4G_5} \cdot \ell^2_3(l_1 + l_2) e^{-\delta_3} - k \Big)^2 
 = - \frac{1}{4}k \Big( \frac{1}{\ell_3} \cdot (l_1 + l_2) e^{-\delta_3}- 1 \Big)^2 ~.
\cr
&=&  -  \frac{1}{4}k \Big(  \frac{2\Phi_L}{1+\Omega} - 1\Big)^2~.
\end{eqnarray}
We freely take $R_6 = \ell_3=1$ in thermodynamic formulae. Comparison of the thermodynamic expressions gives the BPS condition: 
\begin{equation}
\label{eqn:BPScond1}
( 1+\Omega)  - 2\Phi_L =  \frac{2\pi i \ell_3}{\beta_H} ~.
\end{equation}
This form of the BPS condition in the grand canonical ensemble is reminiscent of its analogue in higher dimensions. \cite{Hosseini:2017mds,Cabo-Bizet:2018ehj,Choi:2018hmj}. 

Comparing instead the supergravity forms of \eqref{eqn:hLsugra} and \eqref{eqn:jLsugra},
we find the BPS condition
\begin{equation}
\label{eqn:BPSv4}
\frac{1}{\ell^2_3}\cdot 2m e^{-2\delta_3} = 
 \frac{1}{\ell_3} \cdot (l_1 + l_2) e^{-\delta_3}- 1 ~.
\end{equation}
The computation leading to this result does not involve taking the square root, so there are no undetermined signs or surprising factors of $i$. Indeed, the BPS condition \eqref{eqn:BPSv4} does not involve any explicit complex number. 

On the other hand, to satisfy the BPS condition with all parameters real, the mass parameter $m$ must be negative in the special case $l_1=l_2=0$
where there is no rotation. This reflects the familiar fact that global AdS$_3$ is BPS and formally equivalent to a BTZ black hole with negative mass $M\ell_3=-1$. It is unsurprising that increased rotation, corresponding to larger $l_1+l_2$, increases the BPS mass. However, the familiar reason for this expectation is spectral flow and that gives dependence that is quadratic in rotation. Ultimately these intuitions are inadequate for our purposes. It is essential that BPS equality \eqref{eqn:BPSv4} applies even if $m$ and $l_{1,2}$ are complex. 

It is instructive to rewrite the BPS condition imposed on the potentials \eqref{eqn:BPScond1} in supergravity variables. Rewriting the left and right hand sides of the equation independently, and dividing both with $1+\Omega$, we find the BPS condition: 
\begin{equation}
    \label{eqn:BPSeq7}
\frac{i}{\ell_3} e^{-\delta_3} \sqrt{2m - (l_1 + l_2)^2}  = \frac{l_1 + l_2}{\ell_3} e^{-\delta_3} - 1 ~.
\end{equation}
Taking the square, we recover the condition \eqref{eqn:BPSv4} on the parameters that appear in the solution. That arguably means the equation is the square-root of \eqref{eqn:BPSv4} but this is not a case where an algebraic equation is a perfect square, since the supposedly ``linear" equation \eqref{eqn:BPSeq7} itself involves a square-root. It is surprising that, in supergravity variables, the BPS condition formulated in terms of energy balance is simpler than the version involving potentials.

\subsection{BPS  Strings with Finite Temperature}
\label{subsec:fin}
Our final result for the 6D black string \eqref{eqn:gen6Dfinal} is presented in terms of the four parameters $(m,l_1,l_2,\delta_3)$. The BPS condition \eqref{eqn:BPSv4} expresses $m$ in terms of of the other three variables so we can parametrize the full family of BPS solutions by $(l_1, l_2,\delta_3)$.

To carry this out explicitly, we note that \eqref{eqn:gen6Dfinal} already gives the rotating $S^3$ in terms $(l_1, l_2,\delta_3)$. The parameters of the AdS$_3$ part of the geometry are packaged as the coordinate positions of the horizon $\bar{r}_\pm$ and those are given in \eqref{eqn:rpmtemp}. For the lower sign, the equation simplifies nicely:
$$
(\bar r_+ - \bar r_- )^2 =  \Big(2m - (l_1 +l_2)^2\Big)e^{-2\delta_3}= - \Big(  (l_1 +l_2) e^{-\delta_3} - \ell_3\Big)^2~.
$$
We note again that, when all variables are real, we must impose {\it two} conditions. One of them is extremality $\bar r_+ = \bar r_-$ and the other sets $(l_1 + l_2) e^{-\delta_3}=1$. According to \eqref{eqn:AdSvariables} this is equivalent to $j_L = k$.
Unfortunately, for \eqref{eqn:rpmtemp} with upper sign, the BPS condition does not simplify the formula significantly:  
$$
(\bar r_+ + \bar r_- )^2 =
\Big( 2\ell_3(l_1+l_2)e^{\delta_3} - \ell_3^2 e^{2\delta_3} - (l_1 - l_2)^2\Big)e^{2\delta_3}
= 4l_1 l_2 e^{2\delta_3} - \Big((l_1 + l_2)e^{-\delta_3} -\ell_3\Big)^2 e^{4\delta_3}~.  
$$

An alternative strategy the leads to explicit solutions recasts the general form of BTZ$\times S^3$ in the canonical form: 
\begin{eqnarray}
\label{eqn:final6D4}
ds^2  & = & \frac{2\pi^2\ell^2_3}{\beta_R\beta_L} \cosh \frac{2\rho}{\ell_3}(-dt^2 + dz^2) +  \frac{\pi^2\ell^2_3}{\beta^2_R}(dt - dz)^2 +  \frac{\pi^2\ell^2_3}{\beta^2_L} (dt + dz)^2  + 
d\rho^2\cr
 && +  \frac{1}{4}\ell^2_3 \Bigg(d\theta^2 + \sin^2\theta \Big(d\psi - \frac{2\Phi_R}{1-\Omega}(dt-dz)\Big)^2  \cr
 &&\quad \quad + 
\Big[ d\phi + \cos\theta \Big( d\psi  - \frac{2\Phi_R}{1-\Omega}(dt-dz)\Big)  - \frac{2\Phi_L}{1 + \Omega}  (dt+dz)\Big]^2\Bigg)~,
\end{eqnarray}
where
\begin{eqnarray}
\label{eqn:PhiRLbeta}
\frac{2\Phi_{R,L}}{1\mp\Omega} & =&  \frac{l_1 \mp l_2}{\ell^2_3} e^{\pm\delta_3}~,\cr
\beta_{R,L} & = &  \frac{2\pi\ell^2_3 e^{\mp\delta_3}}{\sqrt{2m - (l_1\mp l_2)^2}}~.
\end{eqnarray}
The BPS condition \eqref{eqn:BPScond1} relates the potentials as 
\begin{equation}
\label{eqn:BPScond8}
\frac{2\Phi_L}{1+\Omega} = 1 + \frac{2\pi i\ell_3}{\beta_L}~.
\end{equation}
Once we allow complex variables, as we must, there is a wide range of possibilities. However, there is a conservative option that is sufficient. First of all, we assume that the AdS$_3$ coordinates $(\tau=it,\bar r, z)$ and $S^3$ coordinates $(\theta, \phi, \psi)$ are real. Moreover, we interpret $\beta_R$ and $\frac{\Phi_R}{1+\Omega}$ as physical variables that characterize the BPS black hole, so those are real as well. That leaves $\beta_L$ and $\frac{\Phi_L}{1-\Omega}$, related through the constraint \eqref{eqn:BPScond8}. They serve as regulators so there is no firm physical motivation for imposing a reality condition. On the other hand, it is preferred that the dipole radius $a$ given by \eqref{eqn:EbetaRL2} is real, if possible, and that requires $\beta_L$ real as well. That leaves $\frac{\Phi_L}{1+\Omega}$ that must be complex and taking the form 
$$
\frac{2\Phi_L}{1+\Omega} = 1 + i \gamma~,\quad \gamma\in\mathbb{R}~. 
$$
This is our conservative proposal. 

The conservative proposal is simple when put in terms of the physical potentials but it is not so simple in the microcanonical ensemble where $E,J, Q_{L,R}$ are given, or in representation employing parameters $(m, l_1 , l_2, \delta_3$.  

For the latter \eqref{eqn:PhiRLbeta} shows 
$$
(l_1 + l_2) e^{-\delta_3}= 1 + i \gamma~,\quad \gamma\in\mathbb{R}~.
$$
but $(l_1 - l_2) e^{\delta_3}$ is real. Additionally, $2m e^{2\delta_3}$ is real, but $2me^{-2\delta_3} = (1 + i \gamma)^2$. This does not appear illuminating. 

In the microcanonical ensemble, the chiral energy $\frac{1}{2}(E+J)$ and the R-charge $Q_R$ are both real, but the corresponding variables in the L-sector
\begin{eqnarray}\label{bpsrelation}
\frac{1}{2}(E-J) &=& \frac{k\pi^2}{\beta^2_L} + k \left( \frac{\Phi_L}{1+\Omega}\right)^2~,
\cr
Q_L &=& \frac{2k\Phi_L}{1+\Omega}~.
\end{eqnarray}
are not. BPS saturation is the linear relation 
$$
E-J = Q_L + \frac{1}{2}k~,
$$
that is satisfied, but with complex numbers on both sides. Therefore, it is not possible to interpret the BPS condition as inequality. 

\subsection{On-shell Action and Indicial Thermodynamics}

In the macroscopic limit, the definition of the partition function gives
\begin{equation}
\label{eqn:partfctdef}
\ln Z = S - \beta (E - \Omega J - \Phi_L Q_L- \Phi_R Q_R) ~.
\end{equation}
The relations \eqref{eqn:EJpot} between $\frac{1}{2}(M\pm J)$ and the potentials give
\begin{eqnarray}
\beta(E-\Omega J) 
& =&  \frac{1}{2} \Big(\beta_R (E+J) +\beta_L (E-J) \Big)~.
\cr
& = & \frac{k\pi^2}{\beta_R} + k \beta_R \left(\frac{\Phi_R}{1-\Omega}\right)^2 + 
\frac{k\pi^2}{\beta_L} + k \beta_L \left(\frac{\Phi_L}{1+\Omega}\right)^2~,
\end{eqnarray}
and the relations \eqref{eqn:QLRpot} for the charges similarly give
$$
\beta \Phi_R Q_R = \frac{2k\Phi^2_R}{1-\Omega}~.
$$
Combining with the entropy \eqref{eqn:bhent2}, 
the partition function \eqref{eqn:partfctdef} becomes:
$$
\ln Z = \frac{k}{\beta_R}\Big( \pi^2 + \beta^2 \Phi^2_R\Big) + 
\frac{k}{\beta_L}\Big( \pi^2 + \beta^2 \Phi^2_L\Big)
$$
This is standard high temperature behavior for any CFT$_2$. For example, it follows by applying modular invariance in the manner done by Cardy. 

The BPS mass relation \eqref{bpsrelation} corresponds to $\Omega=1$ and $\Phi_L=1$. We highlight BPS states by defining variables $\Omega' = \beta(\Omega -1) = -\beta_R$, $\Phi'_L = \beta(\Phi_L -1)$, and $\Phi'_R=\beta\Phi_R$. The primes would have denoted derivative with respect to the temperature $T=\beta^{-1}$ in the limit $\beta\to\infty$, but here we keep $\beta$ finite. Note that $\beta_R = \beta(1-\Omega)=-\Omega'$ and $\beta_L = \beta(1+\Omega)= \Omega' + 2\beta$. With the primed notation that is adapted to the BPS limit, the constraint \eqref{eqn:BPScond1} between the potentials becomes
\begin{equation}
\label{eqn:BPSconstra}
2\Phi'_L - \Omega' = 2\pi i~.
\end{equation}

The superconformal index is defined only for potentials that satisfy the constraint \eqref{eqn:BPSconstra}. It is $\ln I = \ln Z + \beta E_{\rm SUSY}$ where the supersymmetric Casimir energy is $E_{\rm SUSY} = - \frac{1}{2} k$. \cite{Assel:2015nca,ArabiArdehali:2018mil}. In the primed variables, we find
\begin{eqnarray}
\label{eqn:indfreeen}
\ln I & = & 
- \frac{1}{2}k \beta - \frac{k}{\Omega'}\Big( \pi^2 + (\Phi'_R)^2\Big) + 
\frac{k}{\Omega'+2\beta}\Big( \pi^2 + (\Phi'_L + \beta)^2 \Big)
\cr
& = & - \frac{1}{2}k \beta - \frac{k}{\Omega'}\Big( \pi^2 + (\Phi'_R)^2\Big) + 
\frac{k}{\Omega'+2\beta}\Big( \pi^2 + (\frac{1}{2}\Omega' +\beta + \pi i)^2 \Big)
\cr
& = & - \frac{1}{2}k \beta- \frac{k}{\Omega'}\Big( \pi^2 + (\Phi'_R)^2\Big) + 
\frac{1}{4}k(\Omega' +2\beta + 4\pi i )
\cr
& = & \frac{k}{\Omega'}\Big( (\Phi'_L)^2 -(\Phi'_R)^2\Big)
\cr
& = & 4k\frac{\Phi'_1\Phi'_2}{\Omega'}~.
\end{eqnarray}
No limit was taken on the variable $\beta$; it simply drops out of the index, as it should. The index depends on only two independent variables, because of the constraint \eqref{eqn:BPSconstra}. In the final step we introduced the variables $\Phi'_{1,2}=\frac{1}{2}(\Phi'_L \mp \Phi'_R)$. That makes it clearer that the indicial free energy \eqref{eqn:indfreeen} is the AdS$_3\times S^3$ analogue of the HHZ potential that is familiar in higher dimensional AdS. 
\section{Summary and Outlook}
\label{sec:discussion}

The purpose of this article is to develop supersymmetric Euclidean solutions that generalize the standard constructions of black holes with AdS$_3\times S^3$ boundary condition. Our key findings are: 
\begin{itemize}
\item 
In the STU-model we constructed two center solutions with AdS$_3\times S^3$ asymptotics, by augmenting mostly electric $5$-charge solutions with dipoles. The resulting solutions are expressed in terms of all $8$ harmonic functions and appear rather generic. However, we find {\bf all solutions reduce to $S^3$ fibered over BTZ black holes}, just like the construction without dipoles. 
\item
The parameters in the final result fall can be divided in two categories. The R-sector is responsible for the entropy and agrees with the obvious Euclidean continuation of familiar Lorentzian black holes. The BPS conditions restricts the L-sector to a ground state but, for complex solutions, this allows for {\bf a one parameter family of purely imaginary Wilson lines/spectral flows}.  
\item 
The {\bf Casimir energy} plays a starring role. Black holes are excitations of the RR-vacuum where the negative Casimir energy of the true NSNS-vacuum is cancelled by a real Wilson line. The one parameter family of Euclidean black hole follow by an additional Wilson line that is purely imaginary. 
\end{itemize}
These results are consistent between the {\bf two complementary methods} we develop. 

This work represents a foundation for future research in multiple directions: 
\begin{itemize}
\item 
The BPS solutions we consider all have ${\rm Re} J_L = k$. Expectations from higher dimensions suggest that generalizations involves multicomponent phases that combine a BPS black hole with a ``gas" that is BPS by itself \cite{Choi:2024xnv,Choi:2025lck,Larsen:2025jqo}. In Lorentzian AdS$_3\times S^3$ such phases may be interpreted as black holes with a supertube or a black ring \cite{Bena:2011hd}.
\item
To the extent all parameters are allowed to be complex we have constructed a very large family of Euclidean black hole solutions, but it is unclear which ones are physically significant. This presents a good testing ground for general proposals of allowability, like those of KSW \cite{Kontsevich:2021dmb,Witten:2021nzp}, the Lorentzian gravitational path integral \cite{Kolanowski:2026gii}, and the minisuperspace \cite{Mahajan:2025bzo}. Results can be compared with somewhat elaborate thermodynamic stability criteria and the minimal family we identify in subsection \ref{subsec:fin}.
\end{itemize}
We plan to report on these questions in the future.

\acknowledgments
We thank
Alejandra Castro, 
Roberto Emparan, 
Vineeth Krishna,
Joan Simon, and
Amitabh Virmani 
for discussions and comments.

This work was supported by DoE grant DE-SC0007859, the Leinweber Institute for Theoretical Physics, and the Department of Physics at the University of Michigan. 

\appendix

\section{Matter in the Six Dimensional Solution}
\label{app:Matter}
Recall that in six dimensions the bosonic matter of the STU solution can be packaged into a real scalar field \eqref{scalar6D} and an unconstrained gauge 2-form \eqref{6dGaugeform}. For the complexified solution \eqref{eqn:D1D5harm} the scalar has the value
\begin{align}
    e^X= \frac{Q_2}{Q_1}~.
\end{align}
It is independent of position.

In order to find the field strength of the gauge 2 field, one starts with \eqref{6dGaugeform}. To evaluate $\star_5 F^1 $, consider the form of $A^1$ in \eqref{eqn:AIsoln}, which can be rewritten more compactly as
\begin{align}
    A^1&= -\frac{1}{Z_1}(dt+\omega)+ B^1~,\\
    B^1&=\frac{H^1}{H^0}(dy+\chi)
-\xi^1~ ,
\\
*d\xi^1&=dH^1~ .
\end{align}
Then 
\begin{align}
    F^1= -d(\frac{1}{Z_1}) \wedge (dt+\omega) +dB^1 -\frac{1}{Z_1} d\omega
\end{align}
The general form of the 5D metric is 

\begin{align}
    ds_5^2=-Z^{-2}(dt+\omega)^2+Z\,ds_4^2~,
\qquad
Z=(Z_1Z_2Z_3)^{1/3}~.
\end{align}
This implies 

\begin{align}
*_5F^1
=
Z^2 *_4 d(Z_1^{-1})
+
Z^{-1}(dt+\omega)\wedge *_4\!\left(dB^1-Z_1^{-1}d\omega\right)~.
\end{align}
The field strength of the gauge 2-form $B$ for the harmonics \eqref{eqn:D1D5harm} with $h_3=1$ has the final form

\begin{align}
    G= -Q_1 \sin \theta \ d\theta \wedge \left(dy -\frac{P^0Q_0}{2Q_1Q_2 }(dt-dz) \right) \wedge \left( d\phi -ia\frac{P^0}{\Sigma}(dt+dz) \right) + \frac{1}{Q_2}a\sinh \gamma \  dt\wedge dz \wedge d\gamma ~.
\end{align}
The first term in the expression is, up to the constant velocity, precisely the volume form of the fibered $S^3$ geometry seen manifestly in the 6D geometry \eqref{eqn:final6D}, while the second term is, up to a constant, the volume form of the AdS$_3$ part of the metric. 

The lack of symmetry between $Q_1$ and $Q_2$ is a standard feature of the unconstrained three-form field strength. The 6D Hodge dual of the field strength $G$ interchanges the roles of $Q_1$ and $Q_2$, so the (anti-)self dual field strength constructed from the gauge fields will have a (anti-)symmetry between $Q_1$ and $Q_2$.

\bibliographystyle{JHEP}
\bibliography{ref-2}

\end{document}